\documentclass{style}
\usepackage{graphicx}
\usepackage{bm}
\usepackage{amsmath,amssymb}

\begin{document}

\title{Nuclear chiral and magnetic rotation in covariant density functional theory %\\
}
\author{
  Jie Meng \email{mengj@pku.edu.cn} \\
  \it State Key Lab of Nuclear Physics {\rm\&} Technology, School of Physics, \\
  \it Peking University, Beijing 100871, China \\
  Pengwei Zhao \email{pwzhao@pku.edu.cn} \\
  \it Physics Division, Argonne National Laboratory, Argonne, Illinois 60439, USA \\
}


\date{}

\maketitle

\begin{abstract}
Excitations of chiral rotation observed in triaxial nuclei and magnetic and/or antimagnetic rotations seen in near-spherical nuclei have attracted a lot of attention.
Unlike conventional rotation in well-deformed or superdeformed nuclei, here the rotational axis is not necessary coinciding with any principal axis of the nuclear density distribution. Thus, tilted axis cranking is mandatory to describe these excitations  self-consistently in the framework of covariant density functional theory (CDFT).
We will briefly introduce the formalism of tilted axis cranking CDFT and its application for magnetic and antimagnetic rotation phenomena. Configuration-fixed CDFT and its predictions for nuclear chiral configurations and for favorable triaxial deformation parameters are also presented, and the discoveries of the multiple chiral doublets (M$\chi$D) in $^{133}$Ce and $^{103}$Rh are discussed.
\end{abstract}

%%%%%%%%%%%%%%%%%%%%%%%%%%%%%%%%%%%%%%%%%%%%%%%%%%%%%%%%%%%%%%%%%%%%%%%%%%%%%%%%

\section{Introduction}
\label{Sect.01}

The challenge of understanding atomic spectra has opened a new era in atomic physics since 1913, when Niels Bohr suggested that such spectra, e.g., beautiful lines with rainbow colors, were rays emitted by an electron deexciting from one orbit to a lower-energy one, and hence corresponded to transitions between two states of the atom. Atomic spectroscopy thus includes the determination of the states of the atom and of their properties which are determined by the associated wavefunctions.

Similarly, nuclear spectroscopy began in the 1950s with the aim of determining the various states of a nucleus and their properties.
However, the experimental spectra of nuclei exhibit much more complexity than those of atoms because the interactions binding the nucleons are more complicated and the external Coulomb potential producing the spherical shell structure in atoms is missing in nuclei.
On the way to understanding at least the basic features of the observed nuclear spectra, two seemingly contradictory models coexist which are equally successful. On one hand, the liquid drop model describes the nucleus as a whole and the observed nuclear spectra are understood as associated with collective vibrations and/or rotations. On the other, the shell model by Mayer and Jensen {\it et al} treats the nucleus as a composite of individual nucleons and describes the observed nuclear spectra from nucleon excitations.
Both models emphasize either the collective or the individual character but ignore the other.
Combining the single-particle and collective models, for instance, couples one or a few particles to a collective rotor or vibrator, leads to the so-called ``unified model'' introduced by Bohr and Mottelson~\cite{Bohr1953Phys.Rev.316,Bohr1953Phys.Rev.717,Bohr1953Mat.Fys.Medd.Dan.Vid.Selsk.}.

As the book of Bohr and Mottelson~\cite{Bohr1975} demonstrated, ``quantal rotation'' has been a very successful concept in describing rotational spectra of nuclei. The nucleus is considered as a quantum charged liquid drop, which can be deformed as a result of its shell structure.
The deformed nuclear density distribution specifies a space orientation and, thus, provides the rotational degree of freedom. Different rotational bands in a given nucleus correspond to collective rotation originating from different combinations of single nucleon states in the deformed potential. According to quantum mechanics, there is no rotation around the symmetry axis and, therefore, there are no rotational bands in spherical nuclei.

It has been generally accepted that the nuclear high angular momenta  are connected with collective rotation modes associated with stable nuclear deformation~\cite{Bohr1975}. For quadrupole deformed nucleus with axial symmetry, many nucleons may collectively rotates around an
axis perpendicular to the symmetry axis of the nuclear density distribution. The 
corresponding rotational states are usually connected with strong electric quadrupole ($E2$) transitions.
The concept of such collective rotation successfully explains many rotational bands observed. It has also stimulated the discoveries and observation of many exciting phenomena in rapidly rotating nuclei during the past decades, such as backbending~\cite{Johnson1971Phys.Lett.605},
angular momentum alignment~\cite{Stephens1972Nucl.Phys.257,Banerjee1973Nucl.Phys.366},
superdeformation~\cite{Twin1986Phys.Rev.Lett.811}, etc.

In the 1990s, the rotational-like states with significant $M1$ transitions in near-spherical Pb isotopes was observed, which was a surprise at that time.
Experimental investigations of this novel type of collective bands with strong $M1$
and weak $E2$ transitions in near spherical nuclei have attracted a lot of attention (for reviews see Refs.~\cite{Hubel2005Prog.Part.Nucl.Phys.1,Clark2000Annu.Rev.Nucl.Part.Sci.1}).
Due to the weak $E2$ transitions, nuclear deformation should play only a minor role in generating   such rotational states. Instead, it was found that the dynamics of valence nucleons in high-$j$ orbitals is responsible for such rotational structure~\cite{Frauendorf2001Rev.Mod.Phys.463}.

\begin{figure}[ht]
\centering
\vspace{-2mm}
\includegraphics[width=0.45\columnwidth]{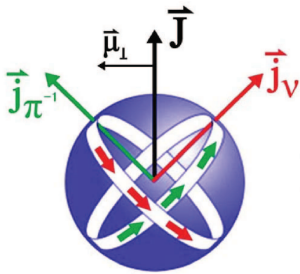}
\vspace{-4mm}
\caption{A schematic illustration for the spin-coupling scheme associated with magnetic rotation. For a near-spherical nucleus, the coupling of the proton-hole $j_{\pi}^{-1}$ and neutron-particle $j_{\nu}$, both in high-$j$ orbitals,
gives the total angular momentum $J$. As a result, a significant transverse component of the magnetic dipole moment $\mu_\perp$ appears and creates strong $M1$ transitions. Taken from Ref.~\cite{Meng2013Front.Phys.55}.
\label{fig1-1}}
\end{figure}

Such bands were explained in terms of the shears mechanism~\cite{Frauendorf1993Nucl.Phys.A259}.
The valence proton holes (particles) and neutron particles (holes) in high-$j$ orbitals generates considerable magnetic moment. This magnetic moment specifies a certain orientation in space and, through its rotation, a series of excited states of rotational character can be built; see Fig.~\ref{fig1-1} for a schematic illustration.
At the bandhead, the neutron particle and proton hole usually tend to align along nearly perpendicular directions to lower the total energy of the system.
When the rotational frequency increases, they start to align towards each other to generate angular momentum, while the total angular momentum barely changes its orientation in the intrinsic frame.
As a result, the so-called ``magnetic rotation'' (MR)~\cite{Frauendorf199452} bands are formed although the density of the system is only weakly deformed.
In magnetic rotation, the order parameter which violates rotational symmetry is the magnetic moment~\cite{Frauendorf1997Z.Phys.A163}. This is quite different from collective rotation in well-deformed nuclei (called electric rotation), where rotational symmetry breaking is induced by the electric quadrupole moments.
However, magnetic rotation in nuclei constitutes an analogy with a ferromagnet in condensed matter, where the isotropy is also violated by the total magnetic moment, i.e., the sum of the atomic dipole moments.
For more detailed discussions, see Ref.~\cite{Meng2013Front.Phys.55}.

Apart from the ferromagnetism, the antiferromagnetism is another very interesting phenomenon in condensed matter.
Unlike the existence of a significant total magnetic moment in a ferromagnet, in an antiferromagnet, the total magnetic moment vanishes.
However, for the latter, all the atomic dipole moments equally align in two opposite directions and, thus, the corresponding isotropy is also violated.

\begin{figure}[ht]
\centering
\vspace{-2mm}
\includegraphics[width=0.5\columnwidth]{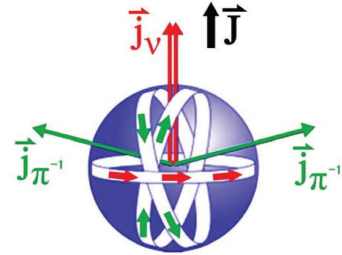}
\vspace{-4mm}
\caption{A schematic illustration for the spin-coupling scheme associated with antimagnetic rotation.
Instead of the shears mechanism for magnetic rotation, the two shears-like configurations here result in a cancellation of the magnetic moments and, thus, no $M1$ transitions appear. Taken from Ref.~\cite{Meng2013Front.Phys.55}.
\label{fig1-2}}
\end{figure}

In an analogy to the antiferromagnet in condensed matter, in nuclear physics, antimagnetic rotation (AMR)~\cite{Frauendorf1996272,Frauendorf2001Rev.Mod.Phys.463} has also attracted a lot of interests in the past decades. Similar to MR, AMR can also be observed in weakly deformed nuclei, and the
valence nucleons in high-$j$ orbitals also play a crucial role in the rotational structure. As depicted in Fig.~\ref{fig1-2}, the high-$j$ neutron particles have their angular momenta aligned in one direction, and the two high-$j$ proton holes align oppositely along the vertical direction. Such an arrangement is usually necessary to achieve the most favored state in energy at the bandhead.
Due to the weak deformation, rotational symmetry is violated by the angular momenta of valence nucleons. Therefore, excitations of rotational character can be built on such a bandhead.
Within this band, the two proton (neutron) blades simultaneously move towards the neutron (proton) angular momentum vector. This is the so-called ``two shears-like mechanism'' which generates the total angular momentum. In this way,  AMR provides a new mode of rotation in weakly deformed nuclei, where valence-nucleon dynamics plays an important role.

Both MR and AMR~\cite{Hubel2005Prog.Part.Nucl.Phys.1,Frauendorf2001Rev.Mod.Phys.463,Clark2000Annu.Rev.Nucl.Part.Sci.1} can happen in a nearly spherical nucleus. Since the early 1990s, a lot of effort has been made to measure MR and AMR bands in experiments. In particular,
the lifetime measurements with high accuracy for four $M1$ bands in $^{198,199}\rm Pb$
performed with Gammasphere provided clear evidences for the shears mechanism~\cite{Clark1997Phys.Rev.Lett.1868}. Since then, a number of MR bands have been observed in the mass regions with $A\sim60$, $A\sim80$, $A\sim110$, $A\sim140$, and $A\sim190$. As summarized in a recent review~\cite{Meng2013Front.Phys.55}, more than 195 magnetic dipole bands have now been observed in 85 nuclei when counting up to the year 2013.

Just like the MR bands, AMR sequences are expected to be observed in the nearly spherical nuclei as well~\cite{Frauendorf2001Rev.Mod.Phys.463}. Therefore, candidates for AMR bands, in principle, are located in the same mass regions as those for MR bands. In fact, it has been demonstrated that the MR and AMR modes can coexist in a single nucleus~\cite{Peng2015Phys.Rev.C44329}.
However, an AMR band differs from a MR one in two aspects. First, no $M1$ transition appears in an AMR band, since the net transverse component of the total magnetic moment is zero. Second, the spin differences between neighboring energy levels in an AMR band are $2\hbar$ rather than $1\hbar$ as in the case in a MR band. This is because an antimagnetic rotor is invariant under a rotation by $\pi$.
Moreover, AMR bands have the feature that the $B(E2)$ values are decreasing with spin, and this has been confirmed by lifetime measurements~\cite{Simons2003Phys.Rev.Lett.162501}.
To date, most AMR bands are observed in the mass regions with $A\sim110$ (see Ref.~\cite{Meng2013Front.Phys.55} and references therein) and, only very recently, have they also been reported in the $A\sim140$ region~\cite{Sugawara2009Phys.Rev.C64321,Rajbanshi2015Phys.Lett.B387}.

Because it had generally been accepted for a long time that nuclear rotation is a collective phenomenon and is associated with substantial deformation, the observation of magnetic rotation brought significant challenges to  nuclear theory. For a full understanding of nuclear rotation, one has to solve, in principle, the time-dependent nuclear many-body problem in a microscopic way, and introduce approximations for simple physics pictures in extreme cases.
Although such microscopic theories are extremely useful, they usually require a great amount of numerical effort. Mean-field theory is one of such widely used approaches for a microscopic description of nuclear structure.
For conventional rotational bands, the cranking mean-field model~\cite{Inglis1956Phys.Rev.1786} is a reliable method since it is a first-order approximation for a variation after projection onto good angular momentum\footnote{Generally, the angular momentum is not a good quantum number in the deformed mean-field approximation  due to the breaking of rotational symmetry. Therefore, the angular momentum projection method is usually employed to restore the violated rotational symmetry and to obtain good angular momentum. A variation after projection approach is, in principle, the most exact method in generating a mean field for nuclear spectroscopy. For more details, see~Ref.~\cite{Ring1980}.}~\cite{Beck1970Z.Phys.26}.

For conventional rotations, the axis in the cranking mean-field model is parallel with the principal axis of the nuclear density distribution with the largest moment of inertia. However, in Ref.~\cite{Frauendorf1993Nucl.Phys.A259}, it was found that a good description of MR bands requires the cranking axis tilted away from the principal axes. 
This leads to the so-called {\it tilted axis cranking} (TAC) model. 
The mean-field solutions of tilted axis rotation were studied in the 1980s~\cite{Kerman1981Nucl.Phys.179,Frisk1987Phys.Lett.B14}. 
However, the physics behind these solutions was missing, and the nuclear deformation was not considered in a self-consistent way in these solutions. 
By using a quadrupole-quadrupole (QQ) Hamiltonian, the TAC model was solved self-consistently for the first time by Frauendorf in 1993~\cite{Frauendorf1993Nucl.Phys.A259}. 
Moreover, it was found that the obtained solutions can account for the $\Delta I = 1$ rotational bands. 
Later on, the quality of the TAC approximation was investigated in comparison with the quantum particle rotor model\footnote{Here, the particle rotor model refers to a quantum version of Bohr and Mottelson's ``unified'' model. Note that a classical particle plus rotor model based on simple angular momentum geometry~\cite{Macchiavelli1998Phys.Rev.C3746,Macchiavelli1998Phys.Rev.C621,Clark2000Annu.Rev.Nucl.Part.Sci.1}, is often used to study the competition between the shears mechanism and collective rotation in MR and AMR bands.} by Frauendorf and Meng~\cite{Frauendorf1996Z.Phys.A263}. 
It was found that the calculated energies and electromagnetic transitions with TAC approach can reproduce the results of the particle rotor model quite well, except near band crossings.

The TAC solutions for triaxial nuclei have attracted considerable attentions in nuclear physics. 
Different from an axial nucleus which has one principal axis as a symmetry axis, none of the three principal axes of a triaxial nucleus is a symmetry axis. 
This indicates that rotational excitations in principle could be built along any principle axes of a triaxial nucleus. 
Usually, a triaxial rigid rotor would favor in energy to rotate around the axis with the largest moment of inertia (MOI).
However, for nuclear systems, besides the ``rigid core'', valence nucleon(s) in high-$j$ orbitals may contribute angular momentum in other two axes regardless of their collective moment of inertia.  
Therefore, one may finally observe a rotation along an arbitrary axis in space for triaxial nuclei.
Such rotation should be described in the framework of cranking mean field as three-dimensional TAC solution 
\footnote{In one-dimensional cranking, the rotational axis is one of the principal axes, as is the case for most conventional rotations. In two-dimensional cranking, the rotational axis is located in a plane defined by two principal axes (principal plane), such as for MR rotation. In three-dimensional cranking, the rotational axis is out of any principal plane, as is the case for chiral rotation.}, which is associated with the exotic phenomenon of nuclear chirality~\cite{Frauendorf1997Nucl.Phys.A131}.

The term ``chiral'' was first used by Lord Kelvin~\cite{Klevin1894}, ``I call any geometrical figure, or group of points, `chiral', and say that it has chirality if its image
in a plane mirror, ideally realized, cannot be brought to coincide with itself''. 
Static chiral symmetries are very common in nature ranging from the macroscopic spirals of snail shells to the microscopic handedness of certain molecules. 
In particular, the spontaneous chiral-symmetry breaking has become one of the most general concepts in modern physics including particle physics, condensed matter physics, and optics.

\begin{figure}[ht]
\centering
\vspace{-2mm}
\includegraphics[width=0.95\columnwidth]{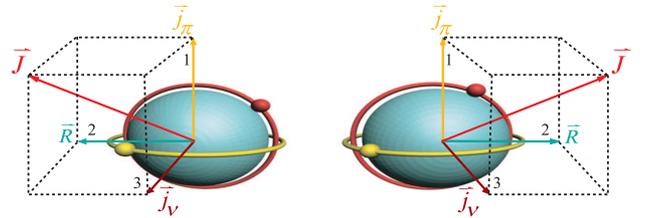}
\vspace{-4mm}
\caption{Left- and right-handed chiral systems for a triaxial odd-odd nucleus in the intrinsic frame. The vector $\vec{J}$ represents the total angular momentum of the nucleus, and $\vec{j_\pi}$, $\vec{j_\nu}$, $\vec{R}$ the angular momenta of the odd-proton, odd-neutron and collective core, respectively. Here the 2 axis indicates the intermediate axis, while the 1 and 3 axes expand the short-long plane of the triaxial nucleus.
 Taken from Ref.~\cite{Meng2010JPhysG.37.64025}.
\label{fig1-3}}
\end{figure}

The existence of spontaneous chiral-symmetry breaking was originally introduced in nuclear physics by Frauendorf and Meng in Ref.~\cite{Frauendorf1997Nucl.Phys.A131}, where they have investigated the rotational excited states in triaxial odd-odd nuclei. 
Fig.~\ref{fig1-3} schematically depicts the chiral geometry in such a nucleus.
The angular momentum of the triaxial core $\vec R$ can align along the intermediate axis since this axis has the largest moment of inertia in the irrotational-like flow. 
However, the angular momenta of the two valence odd nucleons can be mutually perpendicular in the short-long plane, if they are respectively of a particle- and hole-like nature. 
In this way, as depicted in Fig.~\ref{fig1-3}, these three angular momenta form two systems, i.e., the left- and right-handed systems. 
They are different from each other by their intrinsic chirality and, thus, related by the chiral operator $\chi = {\cal T} {\cal R}(\pi)$. 
Here ${\cal T}$ denotes the time reversal and ${\cal R}(\pi)$ spatial rotation by $\pi$. 
The broken chiral symmetry in the intrinsic frame should be restored in the laboratory frame. 
This gives rise to the so-called {\it chiral doublet bands}, which consists a pair of nearly degenerate $\Delta I=1$ bands with the same parity. 

Experimentally, candidate chiral doublet bands have been proposed in a number of odd-odd, odd-$A$ or even-even nuclei in the $A\sim80$, $A\sim100$, $A\sim130$, and $A\sim190$ regions. (see Ref.~\cite{Meng2010JPhysG.37.64025} and references therein).
The best known examples for chiral doublet bands include $^{128}$Cs~\cite{Grodner2006Phys.Rev.Lett.172501},
$^{126}$Cs~\cite{Grodner2011Phys.Lett.B46}, $^{106}$Rh~\cite{Joshi2004Phys.Lett.B135}, $^{135}$Nd~\cite{Zhu2003Phys.Rev.Lett.132501,Mukhopadhyay2007Phys.Rev.Lett.172501}.

The appearance of chiral doublet bands is one of the most important probes for the existence of nuclear triaxiality.
Along this line, the phenomenon of multiple chiral doublet bands (M$\chi$D) introduced in Ref.~\cite{Meng2006Phys.Rev.C37303}, i.e., more than one pair of such bands is presented in a single nucleus, corresponds to triaxial shape coexistence since every pair differs from others in its triaxial deformation and multiparticle configuration. The existence of M$\chi$D was demonstrated for the Rh isotopes by CDFT in Refs.~\cite{Meng2006Phys.Rev.C37303,Peng2008Phys.Rev.C24309,Li2011Phys.Rev.C37301}. The likelihood of chiral bands with different configurations was also discussed in the TAC calculations for the bands observed in $^{105}$Rh~\cite{Timar2004Phys.Lett.B178,Alcantara-Nunez2004Phys.Rev.C24317}.
The first experimental evidence for the existence of M$\chi$D was reported in $^{133}$Ce in 2013~\cite{Ayangeakaa2013Phys.Rev.Lett.172504}. It was found that the negative
parity bands 5 and 6 with a $\pi(1h_{11/2})^2\otimes \nu(1h_{11/2})^{-1}$ configuration
and the positive parity bands 2 and 3 with the $\pi[(1g_{7/2})^{-1}(1h_{11/2})^1] \otimes
\nu(1h_{11/2})^{-1}$ configuration form two pairs of chiral doublet bands. Obviously, these are
two distinct chiral doublet bands due to their different parity. The observation of the M$\chi$D phenomenon also confirms the manifestation of triaxial shape coexistence in this nucleus.

\begin{figure}[ht]
\centering
\vspace{-2mm}
\includegraphics[width=0.7\columnwidth]{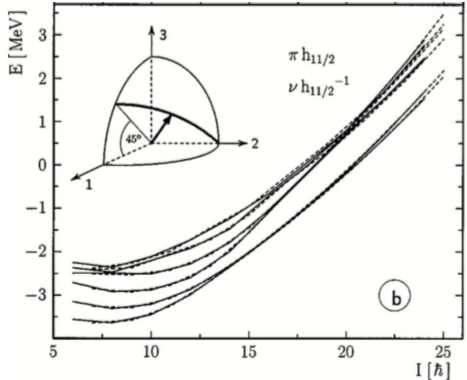}
\vspace{-4mm}
\caption{
Rotational levels of a proton $h_{11/2}$ particle and a neutron $h_{11/2}$ hole coupled to a triaxial rotor with $\gamma=30^\circ$. Solid (dashed) lines correspond to even (odd) spins. The insets show the orientation of the angular momentum, where 1, 2, and 3 correspond to the short, intermediate, and long principal axes, respectively. The angular momentum vector moves along the heavy arc, and the position displayed corresponds to the spin interval $13<I<18$, where the two lowest bands are nearly degenerate. Taken from Ref.~\cite{Frauendorf1997Nucl.Phys.A131}.
\label{fig1-4}}
\end{figure}

Apart from the M$\chi$D with distinct configurations, it is also very interesting to study the robustness of the chiral geometry against the intrinsic excitation.
As an example, in Ref.~\cite{Frauendorf1997Nucl.Phys.A131}, Frauendorf and Meng studied a system of a high-$j$ particle and a high-$j$ hole coupled to a triaxial rotor.
As displayed in Fig.~\ref{fig1-4}, as a consequence of the chiral geometry, the two lowest bands are nearly degenerate within the spin interval $13<I<18$. Moreover, based on the same chiral configuration, one can see that such near degeneracy also occurs for the higher excited bands in certain spin intervals.
This indicates that a chiral geometry may remain in these excited bands as well.
This phenomenon was investigated in more details in several theoretical calculations~\cite{Droste2009Eur.Phys.J.A79,Chen2010Phys.Rev.C67302,Hamamoto2013Phys.Rev.C24327}.
Such predictions have also been confirmed in a recent experiment for $^{103}$Rh~\cite{Kuti2014Phys.Rev.Lett.32501}.

Theoretically, the TAC mean-field model is one of the most successful models for the description of nuclear rotation.
The tilted axis cranking solution and the physics behind it were first introduced in Ref.~\cite{Frauendorf1993Nucl.Phys.A259}, and later discussed thoroughly with the pairing-plus-quadrupole model and the Strutinsky shell-correction model~\cite{Frauendorf2000Nucl.Phys.A115,Frauendorf2001Rev.Mod.Phys.463}. So far, many applications have been performed with these models.
Self-consistent methods based on more realistic two-body interactions are required for a more fundamental investigation, including all important effects such as core polarization and nuclear currents~\cite{Olbratowski2004Phys.Rev.Lett.52501,Zhao2011Phys.Rev.Lett.122501,Meng2013Front.Phys.55}. Such calculations are more challenging, but are feasible in both relativistic~\cite{Madokoro2000Phys.Rev.C61301,Peng2008Phys.Rev.C24313,Zhao2011Phys.Lett.B181} and nonrelativistic~\cite{Olbratowski2004Phys.Rev.Lett.52501,Olbratowski2002ActaPhys.Pol.B389,Olbratowski2006Phys.Rev.C54308} density functional theories (DFTs).

The DFT with a few number of parameters provides a very
successful description of the nuclear properties in both ground and excited states all over the nuclear
chart.Based on the density functionals, rotational excitations have been described by the
principal-axis cranking Hartree-Fock (HF) or Hartree-Fock-Bogoliubov (HFB) approximations with a zero-range Skyrme force~\cite{Fleckner1979Nucl.Phys.A288,Flocard1982Nucl.Phys.A285,Bonche1987Nucl.Phys.A115} and with the density dependent Gogny force~\cite{Egido1993Phys.Rev.Lett.2876} non-relativistically.
A three-dimensional cranking version, which is capable to study nuclear aplanar rotation, has been developed based on the Skyrme density functional and successfully applied  to the magnetic rotation in $^{142}$Gd~\cite{Olbratowski2002ActaPhys.Pol.B389}, and chirality in $A\sim130$ mass regions~\cite{Olbratowski2004Phys.Rev.Lett.52501,Olbratowski2006Phys.Rev.C54308}.

The covariant DFT takes Lorentz symmetry into account
in a self-consistent way and has received wide attention due to
its successful description of a large number of nuclear phenomena in stable as well as exotic
nuclei~\cite{Ring1996Prog.Part.Nucl.Phys.193,Vretenar2005Phys.Rep.101,Meng2006Prog.Part.Nucl.Phys.470}. 
The covariant DFTs with large scalar and vector fields of the order of a few hundred MeV, provide a more efficient
description than the non-relativistic ones which hide these scales. The relativistic framework can lead to reasonable nuclear saturation properties for nuclear
matter with the Brueckner techniques~\cite{Brockmann1990Phys.Rev.C1965,Brockmann1992Phys.Rev.Lett.3408}, reproduce well the isotope
shifts data in the Pb region~\cite{Sharma1993Phys.Lett.B9}, give naturally the spin-orbit potential, explains the origin of pseudospin symmetry~\cite{Arima1969Phys.Lett.B517,Hecht1969Nucl.Phys.A129} as a relativistic symmetry~\cite{Ginocchio2005Phys.Rep.165,Liang2015Phys.Rep.1}, and be reliable for exotic nuclei with extreme proton or neutron number~\cite{Zhao2010Phys.Rev.C54319,Zhao2012Phys.Rev.C64324,Zhao2014Phys.Rev.C11301}, etc.
For nuclear rotation, in particular, covariant density functional theory (CDFT) provides a consistent description of
currents and time-odd fields, and the included nuclear magnetism~\cite{Koepf1989Nucl.Phys.A61} plays an important role in both principal axis cranking rotations~\cite{Konig1993Phys.Rev.Lett.3079,Afanasjev2000Phys.Rev.C31302,Afanasjev2010Phys.Rev.C34329} and tilted axis rotations~\cite{Peng2009Chin.Phys.Lett.32101,Liu2012Sci.ChinaPhys.Mech.Astron.2420}.

Based on the CDFT, Koepf and Ring~\cite{Koepf1989Nucl.Phys.A61,Koepf1990Nucl.Phys.A279} first developed principal axis cranking calculations. K\"{o}nig and Ring~\cite{Konig1993Phys.Rev.Lett.3079} introduced the nuclear currents which are the source of magnetism in the Dirac equation. Nuclear magnetism is found to be crucial for the quantitative understanding of nuclear moments of inertia. Large scale cranking relativistic Hartree-Bogoliubov calculations with the Gogny force D1S in the pairing channel have been performed and summarized in Refs.~\cite{Afanasjev2000Nucl.Phys.A196,Afanasjev2013Phys.Rev.C14320}.

Three dimensional cranking CDFT has been developed in Ref.~\cite{Madokoro2000Phys.Rev.C61301} with the meson-exchange effective interaction. However, it requires a very large numerical resources to perform the calculation. Therefore, the three dimensional cranking CDFT is used only to examine its applicability to  magnetic rotation for relatively light nuclei such as $^{84}\rm Rb$~\cite{Madokoro2000Phys.Rev.C61301}, which require only a two-dimensional cranking calculation.

Focusing on magnetic rotational bands, the two-dimensional cranking CDFT with the  meson-exchange interaction has been established in Ref.~\cite{Peng2008Phys.Rev.C24313}. Compared with three-dimensional cranking CDFT, the two-dimensional cranking CDFT reduced the computing time considerably and, thus, allows its application for medium heavy nucleus such as $^{142}$Gd~\cite{Peng2008Phys.Rev.C24313}.

The point-coupling CDFT has become very popular recently due to the following advantages: 1) it avoids the possible physical constraints introduced by using explicitly the Klein-Gordon equation, in particular the fictitious $\sigma$ meson; 2) the study on the naturalness of the interaction~\cite{Friar1996Phys.Rev.C3085,Manohar1984Nucl.Phys.B189} in effective theories becomes possible for related nuclear structure problems; 3) it is easier to include the Fock terms~\cite{Sulaksono2003Ann.Phys.36,Liang2012Phys.Rev.C21302} and to investigate its relationship to the nonrelativistic ones~\cite{Sulaksono2003Ann.Phys.354}.

In Ref.~\cite{Zhao2011Phys.Lett.B181}, with a point-coupling interaction, the TAC CDFT model was developed which introduces further simplifications and considerably reduces the computing time.
It has been successfully applied for MR ranging from light nuclei $^{60}\rm Ni$~\cite{Zhao2011Phys.Lett.B181} and $^{58}$Fe~\cite{Steppenbeck2012Phys.Rev.C44316} to heavy nuclei $^{198,199}\rm Pb$~\cite{Yu2012Phys.Rev.C24318}. It self-consistently provides also a microscopic investigation for AMR in $^{105}\rm Cd$~\cite{Zhao2011Phys.Rev.Lett.122501,Zhao2012Phys.Rev.C54310}.
Recently, the pairing correlations have also been implemented self-consistently and microscopically by solving the TAC relativistic Hartree Bogoliubov equations~\cite{Zhao2015Phys.Rev.C34319}.

In the following, we mainly focus on microscopic covariant (relativistic) density functional investigations for  tilted axis rotation in nuclei including magnetic rotation, antimagnetic rotation, nuclear chirality and multi chirality.
In Section~\ref{Sect.02}, a sketch of the rotating mean field and the corresponding symmetry breaking
will be introduced.
In Section~\ref{Sect.03}, examples on magnetic rotation, antimagnetic rotation, and nuclear (multi-) chirality will be presented. Finally, a summary and perspectives are given in Section~\ref{Sect.04}.

%%%%%%%%%%%%%%%%%%%%%%%%%%%%%%%%%%%%%%%%%%%%%%%%%%%%%%%%%%%%%%%%%%%%%%%%%%%%%%%%
\section{Rotating Mean Field and Symmetry Breaking}
\label{Sect.02}

\subsection{Concepts related to a mean field}\label{Subsec2.1}

The occurrence of the magic numbers 2, 8, 20, 28, 50, $\ldots$ in nuclei has been one of the strongest evidences for the validity of an average mean field in which the nucleons can move independently~\cite{Mayer1955}. For the nucleons in a nucleus, although there exist a priori no such central field, one can build such an average potential from the interactions among all nucleons in the nucleus. A model with such an average potential treats the nucleons as completely independent moving ones\footnote{Here, the nucleon-nucleon interaction is taken into account in an indirect way, since it gives rise to the average potential.}.
In practical applications, an analytic ansatz for the average potential which is determined by fitting the experimental single-particle levels is widely used for simplicity.
In principle, however, one can derive the form of such an average field from a microscopic two-body nucleon-nucleon force by means of the Hartree-Fock method, for example see Ref.~\cite{Ring1980}. Even starting with a phenomenological nucleon-nucleon interactions, the derived Hartree-Fock potentials are very important to justify the analytic potentials ansatz and to
test different approximations and assumptions adopted.

The Woods-Saxon potential is one of the most successful mean field potential,
\begin{equation}
  V(r,\theta,\phi) = -V_0 \left[ 1 + \exp\left( \frac{r-R(\theta,\phi)}{a(\theta,\phi)}\right)\right]^{-1}.
\end{equation}
Here, one has to allow for a small dependence of $a(\theta,\phi)$ on the angles $\theta$ and $\phi$ in order to get a constant surface diffuseness for deformed nuclei (for more details, see Ref.~\cite{Bohr1975}). Moreover, one needs to consider the spin-orbit term since it plays a very important role for the level structure of nuclei,
\begin{equation}
  V_{LS} = \lambda (\bm{\nabla}V(r,\theta,\phi)\times\bm{p})\bm{s}.
\end{equation}
In general, the involved parameters $V_0$, $\lambda$, as well as the expressions of the parameterized $R(\theta,\phi)$ and $a(\theta,\phi)$, are determined by directly fitting to the experimental single-particles levels.

Another useful approximation to the average potential is the modified harmonic oscillator potential. Such study was originally carried out by Nilsson~\cite{Nilsson1955Mat.Fys.Medd.Dan.Vid.Selsk.1} with the deformed harmonic oscillator Hamiltonian,
\begin{equation}
  h = - \frac{\hbar^2}{2m}\Delta + \frac{m}{2}\omega_\perp^2(x^2+y^2) + \frac{m}{2}\omega_z^2z^2 + C\bm{l}\cdot\bm{s} + D\bm{l}^2,
\end{equation}
where $C$ gives the strength of the spin-orbit force, and $D \cdot \bm{l}^2$ shifts the levels with higher $l$-values downward to achieve better agreement with experimental single-particle spectra\footnote{In order to achieve better agreement with the experimental data, there have been different types of the $\bm{l}^2$ term introduced since the original version was proposed. }.

The models with an analytic average potential work quite successfully in describing the single-particle properties of nuclei. However, in contrast to Hartree-Fock calculations, they fail to correctly reproduce the bulk properties to which all nucleons contribute, such as the total binding energy.
Strutinsky~\cite{Strutinsky1967Nucl.Phys.A420,Strutinsky1968Nucl.Phys.A1} invented a very elegant method, the Strutinsky shell correction approach, in which the quantal description of the phenomenological average potential is
combined with the macroscopic description with the liquid drop model (LDM).
This method is regarded as a major leap forward for the nuclear many-body problem.
It is able to reproduce not only the experimental ground states energies of nuclei, but also their dependence on deformation parameters.

In the Strutinsky method, one calculates the total energy as
\begin{equation}
  E(N, Z) = E_{\rm LDM} + E_{\rm osc},
\end{equation}
where $E_{\rm LDM}$ is a smooth part well represented by the liquid drop model, and $E_{\rm osc}$ is
the quantal shell correction energy due to the occurrence of shell closures,
\begin{equation}
  E_{\rm osc} = \sum\limits_{i=1}^A e_i - {\tilde E}_{\rm sh}
\end{equation}
with $e_i$ the single-particle energy of the state $i$, and ${\tilde E}_{\rm sh}$ the smooth Strutinsky energy.
The Strutinsky method to calculate the quantal part $E_{\rm osc}$ can be found in many textbooks, e.g., in Ref.~\cite{Ring1980,Nilsson2005}.
The applications of the Strutinsky shell correction method in calculating the nuclear masses and/or the energy potential surface with respect to the deformation can be found in many references, for a review see, e.g., Ref.~\cite{Brack1972Rev.Mod.Phys.320}.

Apart from the Strutinsky shell correction model, another model, the {\it pairing-plus-quadrupole model}, has also been widely used due to its simplicity (for a review, see Ref.~\cite{Bes1969Adv.Nucl.Phys.129}).
In this model, one assumes a simple pairing-plus-quadrupole Hamiltonian which has the form,
\begin{equation}
  H = H_{sph}-\frac{\chi}{2}\sum\limits_{\mu=-2}^2Q^\dagger_\mu Q_\mu-GP^\dagger P,
 \end{equation}
with the quadrupole operator,
\begin{equation}
  Q_\mu = \sum\limits_{kk'} \langle k| r^2Y_{2\mu}|k'\rangle c_k^\dagger c_{k'},
\end{equation}
and the creation operator for a Cooper pair,
\begin{equation}
  P^\dagger = \sum\limits_{k>0} c_k^\dagger c^\dagger_{\bar k}.
\end{equation}
Here $\bar k$ is the time reversed state of $k$. The strengths of the force $\chi$ and $G$ depend on the configuration space under consideration and are determined by the experimental data. With such a pairing-plus-quadrupole Hamiltonian, one can prove that the solution of the resulting Hartree-Fock-Bogoliubov (HFB) equations just corresponds to a Nilsson diagonalization with variable deformation parameters related to $\chi$, a subsequent BCS calculation with constant gap parameter related to $G$, and a minimization of the total energy (for more details, see Ref.~\cite{Ring1980}).

The pairing-plus-quadrupole model explains the Nilsson model with BCS approximation very well.
However, it is only a model which contains several key and gross features of a realistic nucleon-nucleon interaction.
More realistic and fundamental approach to derive a self-consistent mean field from a microscopic nucleon-nucleon force is highly demanded.

\subsection{Spontaneous symmetry breaking}\label{Subsec2.2}

The spontaneous symmetry breaking is one of the most basic features of the self-consistent mean-field method. In the relativistic case, so far, most covariant density functional theories are carried out with a mean-field approximation and, thus, the spontaneous symmetry breaking also plays a very important role. In the following, the spontaneous symmetry breaking in the covariant density functional theory will be explained.

The classic version of the covariant density functional theory is the relativistic mean field (RMF) model developed by Walecka and Serot in 1970s~\cite{Serot1986Adv.Nucl.Phys.1,Serot1997Int.J.Mod.Phys.E515}.
In the earlier RMF models, a nucleus is regarded as a composition of a certain number of Dirac nucleons. The interaction between two nucleons is conveyed by exchanging mesons. At the very beginning, only the scalar $\sigma$ meson and vector $\omega$ meson are considered.
Such a simple scheme has already been able to describe several important features of nuclear structure, e.g., nuclear saturation and strong spin-orbit interaction.
Since then, RMF models have been further improved in many aspects by many groups, and also been applied to describe many properties of both stable and exotic nuclei~\cite{Ring1996Prog.Part.Nucl.Phys.193,Vretenar2005Phys.Rep.101,Meng2006Prog.Part.Nucl.Phys.470}.
For a satisfied description of nuclear bulk and single-particle properties, the isovector-vector $\rho$ meson was included in addition to the $\sigma$ and $\omega$ mesons.
Moreover, nonlinear meson self-interactions, which are associated with medium dependences of the effective interactions, were found to be important to achieve a quantitative description of nuclear matter and finite nuclei.
An alternative way to include such medium dependences is to explicitly assume that the meson-nucleon couplings changes with the nucleon densities. More details on the meson-exchange versions of CDFT can be found in Refs.~\cite{Ring1996Prog.Part.Nucl.Phys.193,Vretenar2005Phys.Rep.101,Meng2006Prog.Part.Nucl.Phys.470}.

The successful RMF models can also been reformulated in the framework of Kohn-Sham density functional theory, which is widely used in the treatment of many-body problems, in particular for Coulombic systems. In the framework of density functional theory, all the observables in principle can be calculated through nucleon density.
Therefore, it is very convenient to build the relativistic energy density functionals based on a point-coupling interaction~\cite{Nikolaus1992Phys.Rev.C1757,Burvenich2002Phys.Rev.C44308,Zhao2010Phys.Rev.C54319} by eliminating the meson degrees of freedom.
Specifically, one can replace an exchanged meson by a contact interaction between nucleons in every channel of the interaction, e.g., the scalar, vector, and isovector channels.
Nowadays, these point-coupling models have attracted more and more attentions.

The starting point of the point-coupling CDFT, in its standard version, is an effective Lagrangian:
 \begin{eqnarray}\label{EQ:LAG}
  {\cal L}&=&{\cal L}^{\rm free}+{\cal L}^{\rm 4f}+{\cal L}^{\rm hot}+{\cal L}^{\rm der}+{\cal L}^{\rm em} \nonumber\\
          &=&\bar\psi(i\gamma_\mu\partial^\mu-m)\psi \nonumber\\
          &&-\frac{1}{2}\alpha_S(\bar\psi\psi)(\bar\psi\psi)
            -\frac{1}{2}\alpha_V(\bar\psi\gamma_\mu\psi)(\bar\psi\gamma^\mu\psi) \nonumber \\
          &&-\frac{1}{2}\alpha_{TS}(\bar\psi\vec{\tau}\psi)(\bar\psi\vec{\tau}\psi)
            -\frac{1}{2}\alpha_{TV}(\bar\psi\vec{\tau}\gamma_\mu\psi)(\bar\psi\vec{\tau}\gamma^\mu\psi) \nonumber\\
          &&-\frac{1}{3}\beta_S(\bar\psi\psi)^3-\frac{1}{4}\gamma_S(\bar\psi\psi)^4
            -\frac{1}{4}\gamma_V[(\bar\psi\gamma_\mu\psi)(\bar\psi\gamma^\mu\psi)]^2 \nonumber \\
          &&-\frac{1}{2}\delta_S\partial_\nu(\bar\psi\psi)\partial^\nu(\bar\psi\psi)
            -\frac{1}{2}\delta_V\partial_\nu(\bar\psi\gamma_\mu\psi)\partial^\nu(\bar\psi\gamma^\mu\psi) \nonumber\\
          &&-\frac{1}{2}\delta_{TS}\partial_\nu(\bar\psi\vec\tau\psi)\partial^\nu(\bar\psi\vec\tau\psi)\nonumber \\
          &&-\frac{1}{2}\delta_{TV}\partial_\nu(\bar\psi\vec\tau\gamma_\mu\psi)\partial^\nu(\bar\psi\vec\tau\gamma_\mu\psi)\nonumber\\
         &&-\frac{1}{4}F^{\mu\nu}F_{\mu\nu}-e\frac{1-\tau_3}{2}\bar\psi\gamma^\mu\psi A_\mu,
  \end{eqnarray}
where $m$ is the nucleon mass, $\psi$ is Dirac spinor field of nucleon, $\vec{\tau}$ is the isospin Pauli matrix,
and $e$ is the charge unit for protons. In addition, $A_\mu$ and $F_{\mu\nu}$ are respectively the four-vector potential and field strength tensor of the electromagnetic field. Here, the arrows represent the vectors in the isospin space and the bold type the space vectors. Greek indices $\mu$ and $\nu$ are the Minkowski indices $0$, $1$, $2$, and $3$.

The Lagrangian contains the free nucleon terms ${\cal L}^{\rm free}$, the four-fermion point-coupling terms ${\cal L}^{\rm 4f}$, the higher order terms ${\cal L}^{\rm hot}$, the derivative terms ${\cal L}^{\rm der}$, and the electromagnetic interaction terms ${\cal L}^{\rm em}$. Here, the medium effects are included through the higher order terms which, in the mean field approximation, also lead to density dependent coupling constants of polynomial form.
The derivative terms are employed to simulate the finite-range effects which are important to describe observables relevant to nuclear density distributions, e.g., nuclear radii.

For the Lagrangian in Eq.~(\ref{EQ:LAG}), there are totally 11 coupling constants, $\alpha_S$, $\alpha_V$, $\alpha_{TS}$, $\alpha_{TV}$, $\beta_S$, $\gamma_S$, $\gamma_V$, $\delta_S$, $\delta_V$, $\delta_{TS}$, and $\delta_{TV}$. 
Here, $\alpha$ refers to the four-fermion term, 
$\beta$ the third-order terms, 
$\gamma$ the fourth-order terms, 
and $\delta$ the derivative couplings.
Moreover, we use $S$, $V$, and $T$ to represent the scalar, vector, and isovector couplings, respectively. The terms associated with $\alpha_{TS}$ and $\delta_{TS}$ represent the isovector-scalar channel. They are neglected in many applications because they play a minor role in improving the description of ground-state properties~\cite{Burvenich2002Phys.Rev.C44308}.
Furthermore, the pseudoscalar $\gamma_5$ and pseudovector $\gamma_5\gamma_\mu$ channels are also neglected in Eq.~(\ref{EQ:LAG}) because their contribution vanishes at the Hartree level to keep the parity conservation.

The Hamiltonian, i.e., the $00$ components of the energy-momentum tensor can be obtained by the Legendre transformation
  \begin{equation}
   H = T^{00} = \frac{\partial{\cal L}}{\partial\dot\phi_i}\dot\phi_i - {\cal L},
  \end{equation}
  where $\phi_i$ represents the nucleon or photon field.

In principle, this many-body Hamiltonian should be invariant under a number of symmetry operations, such as translation, rotation, plane reflection, time reversal, etc. In the framework of density functional theory, however, it is not that the exact eigenstates are calculated, but instead, one looks for an approximation of the eigenstates with simple wavefunctions such as Slater determinants.
Therefore, the nuclear ground-state wavefunction $\vert\Phi\rangle$ is assumed as a Slater determinant
  \begin{equation}\label{IntrinsicWF}
   \vert\Phi\rangle=\prod\limits_{k=1}^Aa^\dag_k\vert-\rangle,
  \end{equation}
where $\vert-\rangle$ is the physical vacuum. Accordingly, the nuclear energy density functional can be obtained by
\begin{equation}
  E_{\rm CDF} = \langle\Phi\vert H\vert\Phi\rangle
\end{equation}
in terms of various densities and currents
  \begin{subequations}\label{currents}
   \begin{equation}
    \rho_S(x)=\langle \Phi\vert :\bar\psi \psi:\vert\Phi\rangle,
   \end{equation}
   \begin{equation}
    j^\mu(x)=\langle \Phi\vert:\bar\psi\gamma^\mu \psi:\vert\Phi\rangle,
   \end{equation}
   \begin{equation}
    \vec j^{\mu}_{TV}(x)=\langle \Phi\vert:\bar\psi\gamma^\mu\vec\tau \psi:\vert\Phi\rangle,
   \end{equation}
  \end{subequations}

With the standard variational approach,  the Dirac equation for a single nucleon can be obtained,
  \begin{equation}\label{Dirac}
     h_D\psi_k=\varepsilon_k\psi_k.
  \end{equation}
Here, the single-particle Hamiltonian $h_D$ reads
\begin{equation}
  h_D = \bm{\alpha}\cdot(-i\bm{\nabla}-\bm{V})+ V +\beta(m+S),
\end{equation}
where the scalar $S(\bm{r})$ and vector $V^\mu(\bm{r})$ potentials are connected with the various densities and currents in a self-consistent way as (for details, see Refs.~\cite{Zhao2010Phys.Rev.C54319,Meng2013Front.Phys.55})
   \begin{eqnarray}\label{potential}
     S(\bm{r})   &=&\alpha_S\rho_S+\beta_S\rho^2_S+\gamma_S\rho^3_S+\delta_S\triangle\rho_S, \\
     V^\mu(\bm{r})&=&\alpha_Vj^\mu +\gamma_V (j^\mu)^3 +\delta_V\triangle j^\mu + e \frac{1-\tau_3}{2}A^\mu\nonumber \\
     &&+\vec\tau\cdot(\alpha_{TV}\vec j^\mu_{TV}+\delta_{TV}\triangle\vec j^\mu_{TV}).
   \end{eqnarray}

Since these self-consistent potentials depend not only on the original Hamiltonian, but also on the solution which is represented by the densities and currents, they do not necessarily show the same symmetries as the Hamiltonian. The great advantage of this symmetry breaking is the fact that it allows us to takes into account, in an approximate way, many-body correlations without losing the simple picture of independent particles~\cite{Ring1980}.
For example, one can include the strong quadrupole correlations in the exact solution of $H$ by a substantial quadrupole deformation in the simple Slater determinant $\vert\Phi\rangle$. Of course, this, at the same time, introduces the violation of the rotational symmetry.
Therefore, the mean-field solution, corresponding to the minimum energy, usually does not keep symmetries of the many-body Hamiltonian. This is the so-called {\it spontaneous symmetry breaking}.
In some specific cases, of course, some symmetries could also remain in the mean field, and this all depends upon the system and the interaction.
Note that the exact solution are often referred to states in the laboratory frame, which are in principle holds all the symmetries of the many-body Hamiltonian. The spontaneous symmetry breaking corresponds to mean-field solutions, and appears in the intrinsic frame.

There are universal connections between the symmetries in the intrinsic frame and the structure of nuclear rotational band. One can find such connections by projecting the angular momentum and parity of a band out of a mean-field solution with given symmetry broken. The results of such studies have been summarized in Ref.~\cite{Frauendorf2001Rev.Mod.Phys.463}. In particular, three symmetry operations are important for the classification of the nuclear rotational bands, i.e, the space inversion (parity), the rotation by an angle of $\pi$ (signature), and the chiral operation (product of signature and time reversal).

\subsection{Tilted axis cranking CDFT}\label{Subsec2.3}

In cranking models, the angular momentum is not a good quantum number and the assumption of uniform rotation is introduced.
The quantal spectra are described by employing semiclassical approximations to the energy and electromagnetic transitions.
It is necessary to check how good these approximations are in the description of experimental observables. 
It is known that principal axis cranking models are very successful approaches for describing nuclear rotations.
However, for the TAC case, one should keep in mind that the rotational axis is tilted away from the principal axis.
This causes the signature symmetry breaking which is different from a principal axis cranking model, and, thus the signature is not a good quantum number any more.

Therefore, in Ref.~\cite{Frauendorf1996Z.Phys.A263}, the TAC approach was investigated in comparison with the quantum particle rotor model, where the angular momentum dynamics is treated exactly. 
The TAC approach quantitatively accounts for both the energies and the intra band transition rates for the low-lying bands with a few quasi particles coupled to an axial rotor.
Moreover, it describes well the bandhead in most cases.
Inaccuracy occurs for the cases associated with a significant anti-alignment of particle angular momentum with respect to the rotational axis at very low rotational frequency.

As a mean-field theory, the TAC approach allows one to easily describe multi-quasiparticle excitations, and to study the influence of the deformation evolution and/or the pairing effects.
Moreover, it provides transparent pictures of the angular momentum coupling in terms of classical vector diagrams, which are of great help to understand the structure of rotation bands.
The TAC model also has its drawback, for example, it is difficult to describe the gradual onset of signature splitting. The mixing of bands with different quasiparticle configurations is even beyond a standard cranking theory.

Tilted axis cranking CDFT (TAC-CDFT) has been developed based on either the meson-exchange interaction~\cite{Madokoro2000Phys.Rev.C61301,Peng2008Phys.Rev.C24313} or the point-coupling interaction~\cite{Zhao2011Phys.Lett.B181}. The details of the TAC-CDFT approach based on the meson-exchange interaction can be found in Ref.~\cite{Peng2008Phys.Rev.C24313}. In the following, the main focus will be the TAC-CDFT approach based on the point-coupling interaction~\cite{Zhao2011Phys.Lett.B181,Meng2013Front.Phys.55}.

Assuming that a nucleus rotates uniformly with a constant rotational frequency around an axis in the $xz$ plane, one can transform the Lagrangian in Eq.~(\ref{EQ:LAG}) into such a rotating frame,
\begin{equation}
  \bm{\Omega}=(\Omega_x,0,\Omega_z)=(\Omega\cos\theta_\Omega,0,\Omega\sin\theta_\Omega),
\end{equation}
where $\theta_\Omega:=\sphericalangle(\bm{\Omega},\bm{e}_x)$ is the tilted angle between the cranking-axis and the $x$-axis.
The corresponding equation of motion for nucleons can be derived equivalently by either from a special relativistic transformation~\cite{Koepf1989Nucl.Phys.A61}, or from the tetrad formalism in general relativistic theory~\cite{Madokoro1997Phys.Rev.C2934}. 
Thus one can finds
 \begin{equation}\label{Eq.Dirac}
   [\bm{\alpha}\cdot(-i\bm{\nabla}-\bm{V})+\beta(m+S)
    +V-\bm{\Omega}\cdot\hat{\bm{J}}]\psi_k=\epsilon_k\psi_k,
 \end{equation}
where $\hat{\bm{J}}$ represents the total angular momentum, and $\epsilon_k$ is the single-particle Routhians. The relativistic scalar $S(\bm{r})$ and vector $V^\mu(\bm{r})$ potentials read
\begin{subequations}\label{Eq.poten}
 \begin{eqnarray}
   S(\bm{r})&=&\alpha_S\rho_S+\beta_S\rho_S^2+\gamma_S\rho_S^3+\delta_S\triangle\rho_S, \\
   V(\bm{r})&=&\alpha_V\rho_V+\gamma_V\rho_V^3+\delta_V\triangle \rho_V \nonumber \\
            &&+\tau_3\alpha_{TV} \rho_{TV}+\tau_3\delta_{TV}\triangle \rho_{TV}+eA^0, \\
   \bm{V}(\bm{r})&=&\alpha_V \bm{j}_V+\gamma_V(\bm{j}_V)^3+\delta_V\triangle \bm{j}_V \nonumber \\
            &&+\tau_3\alpha_{TV} \bm{j}_{TV}+\tau_3\delta_{TV}\triangle \bm{j}_{TV}+e\bm{A}.
 \end{eqnarray}
\end{subequations}
As usual, it is assumed that the isospin degree of freedom is not mixed in the single-particle states. This means that the single-particle states are the eigenstates of $\tau_3$. Therefore, only the third component of isovector potentials survives. The Coulomb field $A^0(\bm{r})$ follows the Poisson's equation
\begin{equation}
  -\triangle A^0(\bm{r}) = e\rho_c.
\label{Laplace}
\end{equation}
The spatial components of the electromagnetic vector potential $\bm{A}(\bm{r})$ are neglected as their contributions are extremely small~\cite{Koepf1989Nucl.Phys.A61,Koepf1990Nucl.Phys.A279}.

Since the time reversal symmetry is violated by the Coriolis term $\bm{\Omega}\cdot\hat{\bm{J}}$ in equation~(\ref{Eq.Dirac}), the currents of nucleon, which are connected with the spatial components of the vector potential $\bm{V}(\bm{r})$ are induced. The densities and currents in Eqs.~(\ref{Eq.poten}) read
\begin{subequations}
 \begin{eqnarray}\label{crank:currents}
   \rho_S(\bm{r})     &=& \sum_{i=1}^A \bar\psi_i(\bm{r})\psi_i(\bm{r})        , \\
   \rho_V(\bm{r})     &=& \sum_{i=1}^A \psi_i^\dagger(\bm{r})\psi_i(\bm{r})        , \\
   \bm{j}_V(\bm{r})   &=& \sum_{i=1}^A \psi_i^\dagger(\bm{r})\bm{\alpha}\psi_i(\bm{r}) , \\
   \rho_{TV}(\bm{r})  &=& \sum_{i=1}^A \psi_i^\dagger(\bm{r})\tau_3\psi_i(\bm{r})      , \\
   \bm{j}_{TV}(\bm{r})&=& \sum_{i=1}^A \psi_i^\dagger(\bm{r})\bm{\alpha}\tau_3\psi_i(\bm{r})    , \\
   \rho_{c}(\bm{r})   &=& \sum_{i=1}^A \psi_i^\dagger(\bm{r})\frac{1-\tau_3}{2}\psi_i(\bm{r}).
 \end{eqnarray}
\end{subequations}
In the above equations, the ``no-sea'' approximation is adopted, i.e., the contribution from the negative-energy states in the Dirac sea are neglected and the sums run over the particles states in the Fermi sea only.

After solving the equation of motion in a self-consistent way, one can obtain the total energy, which is to be compared with the corresponding experimental data,
\begin{equation}
  \label{Eq.Etot}
   E_{\rm tot} = E_{\rm kin} + E_{\rm int} + E_{\rm cou} + E_{\rm c.m.},
\end{equation}
which is composed of a kinetic part
\begin{equation}
   E_{\rm kin} = \int d^3\bm{r}\sum\limits_{i=1}^A\psi_i^\dagger[\bm{\alpha}\cdot\bm{p}+\beta m]\psi_i,
\end{equation}
an interaction part
\begin{eqnarray}
    E_{\rm int}&=&\int d^3\bm{r} \left\{\frac{1}{2}\alpha_S\rho_S^2+\frac{1}{3}\beta_S\rho_S^3+\frac{1}{4}\gamma_S\rho_S^4
    +\frac{1}{2}\delta_S\rho_S\Delta\rho_S\right.\nonumber \\
    &&+\frac{1}{2}\alpha_V(\rho_V^2-\bm{j}\cdot\bm{j})
    +\frac{1}{2}\alpha_{TV}(\rho_{TV}^2-\bm{j}_{TV}\cdot\bm{j}_{TV})\nonumber \\
    &&+\frac{1}{4}\gamma_V(\rho_V^2-\bm{j}\cdot\bm{j})^2
    +\frac{1}{2}\delta_V(\rho_V\Delta\rho_V-\bm{j}\Delta\bm{j})\nonumber \\
    &&+\left.\frac{1}{2}\delta_{TV}(\rho_{TV}\Delta\rho_{TV}-\bm{j}_{TV}\Delta\bm{j}_{TV})\right\},
\end{eqnarray}
an electromagnetic part
\begin{equation}
    E_{\rm cou} = \int d^3\bm{r}\frac{1}{2}eA_0\rho_c,
\end{equation}
and the center-of-mass (c.m.) correction energy $E_{\rm c.m.}$ which is included for the motion of the center-of-mass. Nowadays, this is usually done by including the microscopic c.m. correction energy~\cite{Bender2000Eur.Phys.J.A467,Zhao2009Chin.Phys.Lett.112102}
  \begin{equation} \label{Eq:Ecm}
   E^{\rm mic}_{\rm c.m.}=-\frac{1}{2mA}\langle\hat{\bm P}^{2}_{\rm c.m.}\rangle,
  \end{equation}
  with $A$ being the mass number and $\hat{\bm{P}}_{\rm c.m.}=\sum_i^A \hat{\bm{p}}_i$ being the total momentum in the c.m. frame. Therefore, the total energy for the nuclear system becomes
  \begin{equation}
    E_{\rm tot} = E_{\rm CDF}+E^{\rm mic}_{\rm c.m.}.
  \end{equation}

For a given rotational frequency $\Omega$, the angular momentum is evaluated through the expectation values of its components $\bm{J}=(J_x,J_y,J_z)$ in the intrinsic frame
\begin{subequations}
\begin{eqnarray}
    J_x&=&\langle\hat{J}_x\rangle=\sum\limits_{i=1}^Aj^{(i)}_x, \\
    J_y&=& 0, \\
    J_z&=&\langle\hat{J}_z\rangle=\sum\limits_{i=1}^Aj^{(i)}_z.
\end{eqnarray}
\end{subequations}
Moreover, the rotational frequency $\Omega$ is related to the corresponding quantum number of angular momentum $I$ by the so-called semiclassical cranking condition for rotational bands,
\begin{equation}
 J = \sqrt{\langle\hat{J}_x\rangle^2+\langle\hat{J}_z\rangle^2}\equiv\sqrt{I(I+1)}.
\end{equation}

For the direction of the angular momentum vector $\bm{J}$, one can define an angle $\theta_J:=\sphericalangle(\bm{J},\bm{e}_x)$ between the angular momentum $\bm{J}$ and the $x$ axis.
In a fully self-consistent calculation, this angle should be determined by requiring that the angular momentum vector is parallel with the angular velocity $\bm{\Omega}$.

The quadrupole moments $Q_{20}$ and $Q_{22}$ are calculated by
\begin{subequations}
\begin{eqnarray}
    Q_{20}&=&\sqrt{\frac{5}{16\pi}}\langle3z^2-r^2\rangle, \\
    Q_{22}&=&\sqrt{\frac{15}{32\pi}}\langle x^2-y^2\rangle,
\end{eqnarray}
\end{subequations}
and the deformation parameters $\beta$ and $\gamma$ can thus be extracted from
\begin{subequations}
\begin{eqnarray}
    \beta&=&\sqrt{a_{20}^2+2a_{22}^2}, \\
   \gamma&=&\arctan\left[\sqrt{2}\frac{a_{22}}{a_{20}}\right],
\end{eqnarray}
\end{subequations}
by using the relations
\begin{subequations}
\begin{eqnarray}
    Q_{20}&=&\frac{3A}{4\pi}R_0^2a_{20}, \\
    Q_{22}&=&\frac{3A}{4\pi}R_0^2a_{22},
\end{eqnarray}
\end{subequations}
with $R_0 = 1.2A^{1/3}~\rm fm$. Here, the sign convention for the definition of $\gamma$ is the same as that in Ref.~\cite{Ring1980}.

The nuclear magnetic moment is calculated in a relativistic way from the nuclear currents
\begin{equation}
   \bm{\mu} = \sum\limits_{i=1}^A\int d^3r\left[\frac{mc^2}{\hbar c}q\psi^\dagger_i(\bm{r})\bm{r}\times\bm{\alpha}\psi_i(\bm{r})+\kappa\psi^\dagger_i(\bm{r})\beta\bm{\Sigma}\psi_i(\bm{r})\right].
\end{equation}
Here, the magnetic moment is in units of the nuclear magneton. The charge $q$ ($q_p=1$ for protons and $q_n=0$ for neutrons) is in units of $e$. Moreover, the free anomalous gyromagnetic ratio $\kappa$ is directly used for the nucleon ($\kappa_p=1.793$ and $\kappa_n=-1.913$).

By using the quadrupole moments and magnetic moment, the transition probabilities $B(M1)$ and $B(E2)$ can be calculated in the semiclassical approximation
\begin{subequations}
\begin{eqnarray}
   B(M1) &=& \frac{3}{8\pi}\mu_{\bot}^2 =\frac{3}{8\pi}(\mu_x\sin\theta_J-\mu_z\cos\theta_J)^2,\\
   B(E2) &=& \frac{3}{8}\left[ Q^p_{20}\cos^2\theta_J+\sqrt{\frac{2}{3}}Q^p_{22}(1+\sin^2\theta_J)\right]^2, \nonumber \\
\end{eqnarray}
\end{subequations}
where $Q^p_{20}$ and $Q^p_{22}$ are the corresponding proton quadrupole moments.

For the numerical solutions of the relativistic tilted axis cranking equations, one can expand the Dirac spinors in three dimensional harmonic oscillator wavefunctions in Cartesian coordinates.
Different from the usual principal axis cranking programs (one-dimensional cranking), where the principal axes of the density distribution are along the $x$, $y$, and $z$ axis, it allows for arbitrary rotation of the density with respect to the intrinsic $y$ axis in the two-dimensional cranking case.
However, the solutions with different rotation along the $y$ axis are essentially degenerate, and numerically this can lead to instabilities in the iterative procedure.
Therefore, the $x$, $y$, and $z$ axes are enforced to be identical with the principal axes of the density distribution by forcing the expectation value $\langle{Q_{2-1}}\rangle=0$ with the quadratic constraint method~\cite{Ring1980}.
Moreover, the rotation bands are built on specific proton and neutron configurations. Therefore, in the calculations, one needs to identify and keep tracing the specific configuration along a band. The details for the numerical calculations can be found in Refs.~\cite{Peng2008Phys.Rev.C24313,Zhao2012Phys.Rev.C54310,Meng2013Front.Phys.55}.

Finally, it is worthwhile to mention that there are no energy density functionals derived directly from realistic nucleon-nucleon interaction so far. Most density functionals currently used are determined by fitting to the ground-state properties for a few nuclei.
However, it turns out that these density functionals generally work quite well for almost all nuclei throughout the nuclear chart.
For example,
the point-coupling density functional PC-PK1 was determined by fitting to observables of 60 selected spherical nuclei, including the binding energies, charge radii, and empirical pairing gaps~\cite{Zhao2010Phys.Rev.C54319}. This functional improves remarkably the description for isospin dependence of binding energies. Its success has been proved in many calculations with CDFT, e.g., the Coulomb displacement energies between mirror nuclei~\cite{Sun2011Sci.ChinaSer.G-Phys.Mech.Astron.210}, fission barriers~\cite{Lu2012Phys.Rev.C11301}, the measured nuclear masses~\cite{Zhang2014Front.Phys.1}, exotic nuclei including their masses~\cite{Zhao2012Phys.Rev.C64324} and quadrupole moments~\cite{Zhao2014Phys.Rev.C11301}, etc.

%%%%%%%%%%%%%%%%%%%%%%%%%%%%%%%%%%%%%%%%%%%%%%%%%%%%%%%%%%%%%%%%%%%%%%%%%%%%%%%%
\section{Selected Examples }
\label{Sect.03}

%%%%%%%%%%%%%%%%%%%%%%%%%%%%%%%%%%%%%%%%%%%%%%%%%%%%%%%%%%%%%%%%%%%%%%%%%%%%%%%%
\subsection{Magnetic Rotation}
\label{Sect.03.01}

The long cascades of $M1$ transitions in Pb nuclei were firstly observed in the early 1990's~\cite{Clark1992Phys.Lett.B247,Baldsiefen1992Phys.Lett.B252,Kuhnert1992Phys.Rev.C133,Clark1993Nucl.Phys.A121,Baldsiefen1994Nucl.Phys.A521}. With improved detector techniques and lots of efforts, lifetime measurements with the Doppler-shift attenuation method (DSAM) were carried out in 1997 for four $M1$-bands in the nuclei $^{198}\rm Pb$ and $^{199}\rm Pb$, and this provided a clear evidence for magnetic rotation~\cite{Clark1997Phys.Rev.Lett.1868}.
Subsequently, another experiment using the recoil distance method (RDM) is preformed for $^{198}\rm Pb$~\cite{Kruecken1998Phys.Rev.C1876}. This experiment provides further support to the existence of the shears mechanism.
Since then many theoretical works have been devoted to these bands and, thus, the $M1$ bands in Pb isotopes become the classic examples for nuclear magnetic rotation.

Magnetic rotation was first investigated with the TAC approach in the framework of the pairing plus quadrupole model. It reproduces the experimental $B(M1)$ values of the dipole bands in $^{198,199}$Pb very well~\cite{Clark1997Phys.Rev.Lett.1868}.
Extensive analyses of the deformation and transition probabilities for the magnetic rotation bands in Pb isotopes have been carried out later in Ref.~\cite{Chmel2007Phys.Rev.C75}.
In addition, others models including the shell model~\cite{Frauendorf1996Nucl.Phys.A41} and the many-particles-plus-rotor model~\cite{Carlsson2006Phys.Rev.C44310} were also used to study the magnetic rotation in the Pb region.
In the following, as an example, the relativistic description for the band 1 in $^{198}$Pb~\cite{Clark1993Nucl.Phys.A121} is shown to demonstrate the relativistic self-consistent description of the MR characteristics. The calculations were carried out with PC-PK1~\cite{Zhao2010Phys.Rev.C54319}, while pairing correlations were not included. More details can be found in Ref.~\cite{Yu2012Phys.Rev.C24318}.

\subsubsection{Single-particle Routhian and configuration}

\begin{figure}[ht]
\centering
\vspace{-2mm}
\includegraphics[width=1.0\columnwidth]{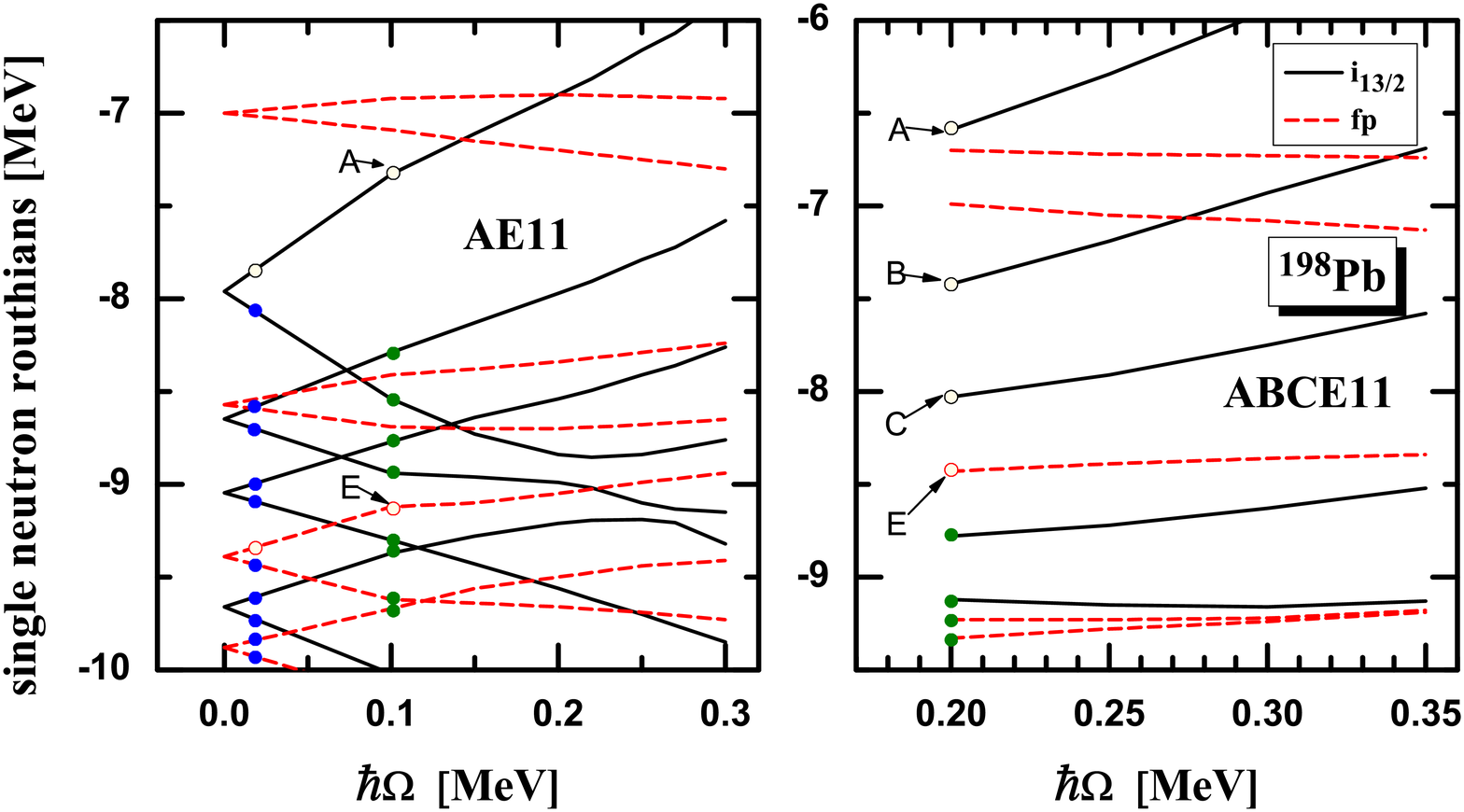}
\vspace{-4mm}
\caption{Single-particle Routhians for the neutrons in $^{198}$Pb with the configurations AE11 and ABCE11. The blue dots indicate the occupied levels at $\Omega=0$ and the green dots indicate the occupied levels at the band heads with the configuration AE11 (left panel) and ABCE11 (right panel). Taken from Ref.~\cite{Yu2012Phys.Rev.C24318}.\label{fig3-1}}
\end{figure}

As suggested in Ref.~\cite{Clark1993Nucl.Phys.A121}, the proton configuration for band 1 in $^{198}$Pb is $\pi[s_{1/2}^{-2}h_{9/2}i_{13/2}]11^-$. 
In Fig.~\ref{fig3-1}, we show the single neutron Routhians in $^{198}$Pb with the configurations AE11 and ABCE11 with respect to the rotational frequency $\Omega$. The short hand notations AE11 and ABCE11 are the same as in Ref.~\cite{Baldsiefen1994Nucl.Phys.A521}, i.e., A, B, C and D denote $\nu i_{13/2}$ holes with positive parity, and E represents a low-$j$ neutron hole in the $fp$ shell ($f_{5/2}$ and $p_{3/2}$) with negative parity. The proton configuration $\pi[s_{1/2}^{-2}h_{9/2}i_{13/2}]11^-$ is abbreviated by its spin number 11.
Therefore, the neutron configuration $\nu [i_{13/2}^{-1}(fp)^{-1}]$ is referred as AE11, $\nu[ i_{13/2}^{-3}(fp)^{-1}]$ as ABCE11, and they correspond to the states before and after the backbending respectively.

In Fig.~\ref{fig3-1}, the $\nu i_{13/2}$ orbitals with positive parity are shown with solid lines and the neutron levels in the $(pf)$ shell with negative parity are indicated by dashed lines.
At a given rotational frequency, the time-reversal symmetry is violated by the Coriolis term. As a consequence, the degenerate time-reversal conjugate states split in energy with the rotational frequency.
The amplitudes of these energy splitting can be several MeV, which mainly depend on the Coriolis term. Due to these energy splittings, the order of the orbitals may be changed with increasing frequency. In such case, one has to fix the configuration by tracing each orbital and keeping their occupation unchanged.

\subsubsection{Energy spectrum}

\begin{figure}[ht]
\centering
\vspace{-2mm}
\includegraphics[width=0.8\columnwidth]{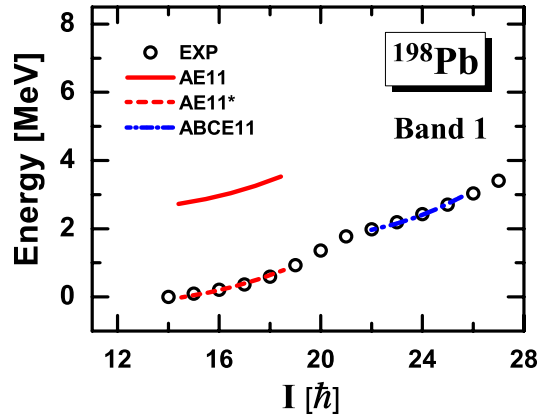}
\vspace{-4mm}
\caption{
Energy spectra in the TAC-CDFT calculations compared with the data~\cite{Gorgen2001Nucl.Phys.A683} for band 1 in $^{198}\rm Pb$. The energies at $I= 22\hbar$ is taken as references for the band 1 in $^{198}\rm Pb$. Energies for the configurations AE11* in $^{198}\rm Pb$ are renormalized to the energies at $I=15\hbar$. Taken from Ref.~\cite{Yu2012Phys.Rev.C24318}, but the energy references are different here.\label{fig3-2}}
\end{figure}

The calculated energy spectrum for band 1 in $^{198}\rm Pb$ is shown in Fig.~\ref{fig3-2} in comparison with the data~\cite{Gorgen2001Nucl.Phys.A683}.
The calculated values are missing when $I=19-21\hbar$ 
since there is no converged solutions.
As discussed in Ref.~\cite{Yu2012Phys.Rev.C24318}, this is due to an occurrences of level crossing in this region which is associated with the backbending phenomenon.
The experimental rotation energies can be reproduced well by the TAC-CDFT calculations, but the calculated bandhead energy of the configuration AE11 is not in agreement with the data.
At the moment, this difference needs to be compensated by shifting the bandhead energy artificially. In comparison with the TAC calculations with pairing plus quadrupole model (PQTAC)~\cite{Chmel2007Phys.Rev.C75}, this disagreement may result from the lack of pairing correlations, since pairing could lower the energy of states with configuration AE11. This needs to be further investigated in the future.

\begin{figure}[ht]
\centering
\vspace{-2mm}
\includegraphics[width=0.8\columnwidth]{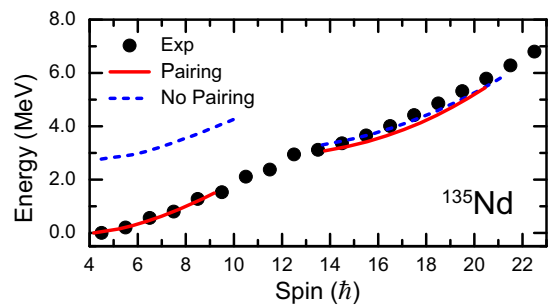}
\vspace{-4mm}
\caption{ Calculated energy spectrum as functions of the angular momentum in comparison with the data of Ref.~\cite{Zhu2003Phys.Rev.Lett.132501} (solid dots). Here, the excitation energies are the relative energy differences with respect to the ground state. Taken from Ref.~\cite{Zhao2015Phys.Rev.C34319}.\label{fig3-3}}
\end{figure}

Although there is no TAC-CDFT calculation with pairing for the MR bands in Pb nuclei so far, but in a very recent work, this has been done for the yrast band in the medium mass nucleus $^{135}$Nd. The calculated energy spectrum is compared with the data~\cite{Zhu2003Phys.Rev.Lett.132501} in Fig.~\ref{fig3-3}. It indicates that the experimental rotational excitation energies for both the one quasiparticle (lower spin part) and three quasiparticle bands (higher spin part) are reproduced well by the self-consistent calculations with pairing. The pairing correlation leads to the correct energy difference between these two configurations. Therefore, the artificial renormalization of the bandhead energies for band 1 in $^{198}\rm Pb$ has been eliminated by including the pairing.

\begin{figure}[ht]
\centering
\vspace{0mm}
\includegraphics[width=0.8\columnwidth]{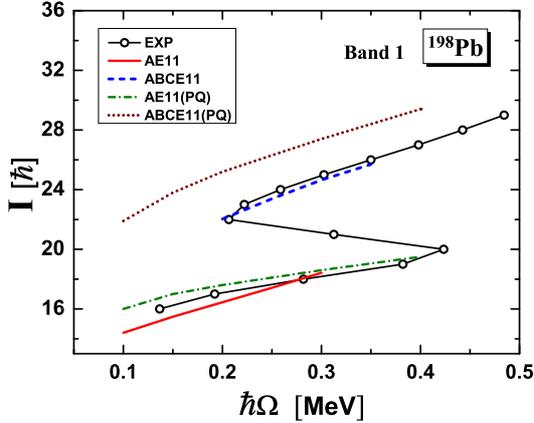}
\vspace{-4mm}
\caption{Angular momenta as  functions of the rotational frequency in the TAC-CDFT calculations
compared with the data~\cite{Gorgen2001Nucl.Phys.A683} and the PQTAC results~\cite{Chmel2007Phys.Rev.C75} for band 1 in $^{198}\rm Pb$. The configurations with ``(PQ)'' denote the corresponding
results of PQTAC calculations. Taken from Ref.~\cite{Yu2012Phys.Rev.C24318}.\label{fig3-4}}
\end{figure}

From the experimental energy spectrum, one can extract the corresponding rotational frequency $\Omega_{\rm exp}$ by the relation
 \begin{equation}
    \hbar\mathit\Omega_{\rm exp}\approx\frac{dE}{dI}=\frac{1}{2}[E_\gamma(I+1\rightarrow I)+E_\gamma(I\rightarrow I-1)].
 \end{equation}
 In Fig.~\ref{fig3-4}, the calculated angular momenta as functions of the rotational frequency for band 1 in $^{198}\rm Pb$ are shown in comparison with the experimental data~\cite{Gorgen2001Nucl.Phys.A683} and the PQTAC results~\cite{Chmel2007Phys.Rev.C75}.
 The data can be well reproduced by both the TAC-CDFT and PQTAC calculations.
 This indicates that the TAC calculations can reproduce the moments of inertia as well.
 The backbending appears clearly in this band, and it is due to the excitation of a pair of neutron holes in the $i_{13/2}$ shell, which corresponds to a configuration transition from AE11 to ABCE11.
 Before the backbending, the differences between the experimental angular momenta and the TAC-CDFT and PQTAC results are lower than $2\hbar$. After the backbending, the angular momenta from PQTAC calculations are higher than the corresponding data by around $3\hbar$, while the TAC-CDFT results agree well with the data.

 It should be mentioned that as addressed in Ref.~\cite{Zhao2015Phys.Rev.C34319}, the superfluidity induced by pairing allows additional mixing in single-particle orbitals, and would influence the total spin in two aspects; i.e., reducing the magnitude of the proton and neutron angular momenta, but expediting the merging of their directions. In the future, it would be interesting to investigate the impact of the pairing correlations on the tilted axis rotation bands in Pb nuclei.

\subsubsection{Shears mechanism}

\begin{figure}[ht]
\centering
\vspace{0mm}
\includegraphics[width=1.0\columnwidth]{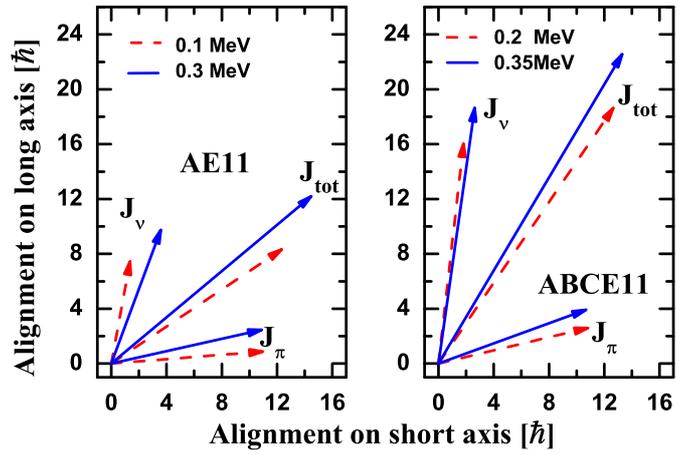}
\vspace{-4mm}
\caption{Composition of the total angular momentum at both the minimum and maximum rotational frequencies in TAC-CDFT calculations for band 1 in $^{198}\rm Pb$. Left (right) panel corresponds to the rotation before (after) backbending. Taken from Ref.~\cite{Yu2012Phys.Rev.C24318}.\label{fig3-5}}
\end{figure}

One of the typical characteristics in magnetic rotation is the shears mechanism.
In Fig.~\ref{fig3-5}, the angular momentum vectors for proton $\bm J_\pi$ and neutron $\bm J_\nu$ as well as the total angular momentum vectors $\bm J_{\rm tot}=\bm J_\pi+\bm J_\nu$ at both the minimum and maximum rotational frequencies are shown. The proton and neutron angular momenta $\bm J_\pi$ and $\bm J_\nu$ are defined as
\begin{equation}
    \bm J_\pi=\langle\bm{\hat J}_\pi\rangle=\sum_{p=1}^{Z}\langle p|\hat J|p\rangle, \quad\quad \bm J_\nu=\langle \bm{\hat J}_\nu\rangle=\sum_{n=1}^{N}\langle n|\hat J|n\rangle,
       \label{eq:AM}
\end{equation}
where the sum runs over all the occupied proton (or neutron) levels.

For this MR band in $^{198}\rm Pb$, we have neutron(s) occupied in the high-$j$ $i_{13/2}$ orbitals, and proton(s) in the $h_{9/2}$ and $i_{13/2}$ orbitals. The angular momenta are mainly from the contribution of these orbitals.
At the band head, the proton particles in the $h_{9/2}$ and $i_{13/2}$ orbitals align the proton angular momentum mainly along the short axis, while the neutron hole(s) in the $i_{13/2}$ shell align the neutron angular momentum mainly along the long axis. The neutron angular momentum become significantly larger after the backbending, because of the breaking of a pair of neutron holes in the $i_{13/2}$ shell. Two blades of a pair of shears are formed by the proton and neutron angular momenta, respectively.
With the increasing rotational frequency, the magnitude of the total angular momentum is increased by aligning these two blades towards each other, while the corresponding direction barely change. This is just in analogy to closing a pair of shears and, thus, the shears mechanism appears clearly.

\subsubsection{Electric and magnetic transitions}

\begin{figure}[ht]
\centering
\vspace{0mm}
\includegraphics[width=0.9\columnwidth]{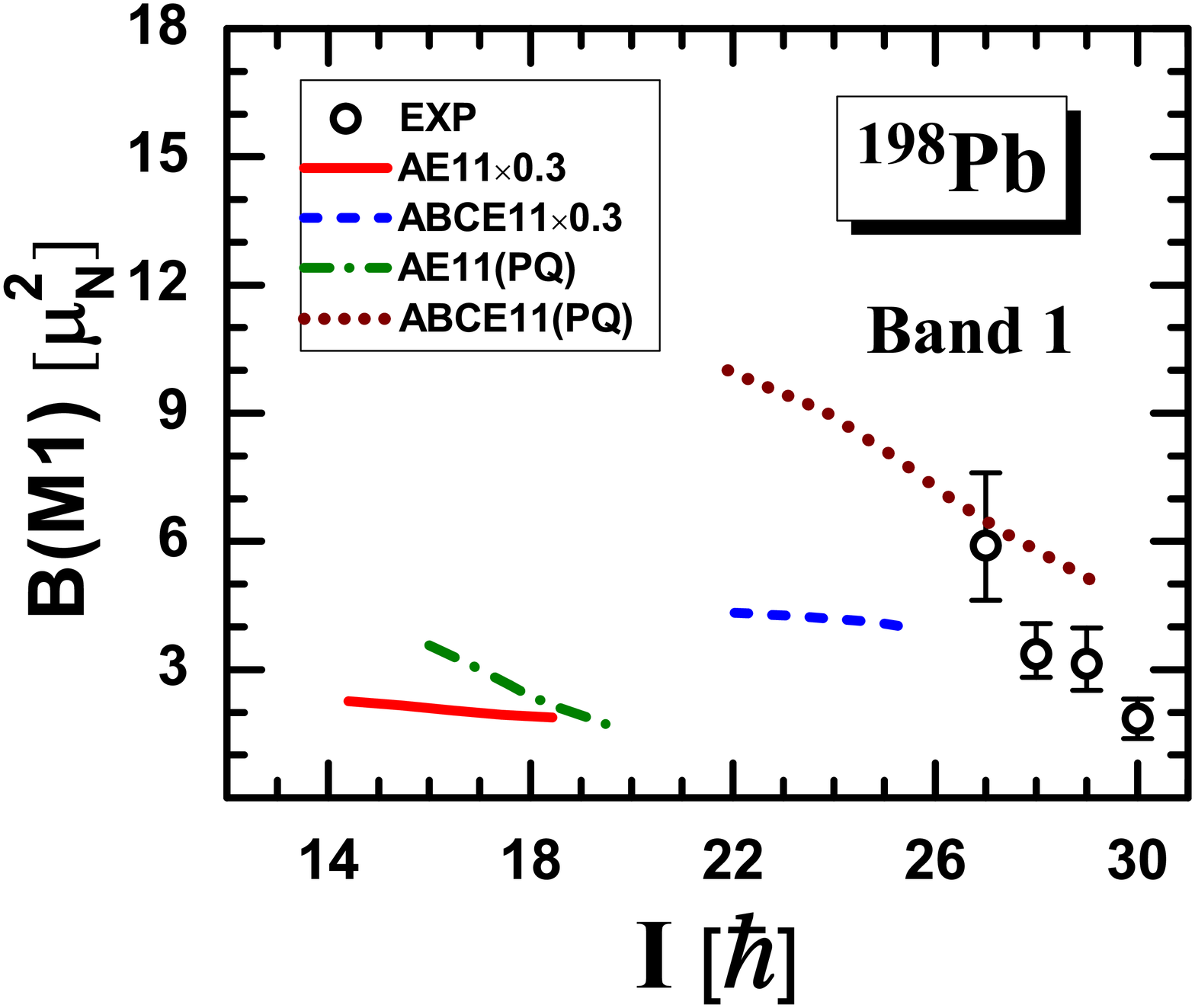}
\vspace{-4mm}
\caption{$B(M1)$ values as  functions of the total angular momentum in the TAC-CDFT calculations compared
with the data and the PQTAC results~\cite{Chmel2007Phys.Rev.C75} for band 1 in $^{198}\rm Pb$. Circles and squares denote experimental data from Ref.~\cite{Clark1997Phys.Rev.Lett.1868} and Ref.~\cite{Kruecken1998Phys.Rev.C1876}, respectively. Taken from Ref.~\cite{Yu2012Phys.Rev.C24318}.\label{fig3-6}}
\end{figure}

The strongly enhanced $M1$ transition probabilities decreasing with the spin is a typical characteristic of magnetic rotation.
In Fig.~\ref{fig3-6}, the $B(M1)$ values from TAC-CDFT~\cite{Yu2012Phys.Rev.C24318} and PQTAC~\cite{Chmel2007Phys.Rev.C75} calculations as well as the corresponding data~\cite{Clark1997Phys.Rev.Lett.1868,Kruecken1998Phys.Rev.C1876} are shown.

The $B(M1)$ values are decreased in the TAC-CDFT calculations with increasing angular momentum.
This is a result of the shears mechanism, and also consistent with the data.
Quantitatively, however, one has to introduce an attenuate factor of 0.3 to describe the $B(M1)$ values.
The microscopic reason of this factor has not been fully understood yet.
On the one hand, it should be noted that the calculated $B(M1)$ values with PQTAC reproduce the experimental data with the standard attenuation factor of 0.6-0.7 for the spin part of the magnetic transition moments.
In the relativistic framework, this corresponds to an attenuation of the anomalous gyromagnetic ratio $\kappa$.
The microscopic reason for this attenuation is a long-standing problem, and in the relativistic case, in fact, not only the anomalous part but also the Dirac part of the magnetic moment need to be attenuated, because the Dirac effective nucleon mass ($M^\ast\sim0.6M$) enhances the relativistic effect on the electromagnetic current~\cite{McNeil1986Phys.Rev.C746}.
A recent work shows that the time-odd fields, one-pion exchange current~\cite{Miyazawa1951Prog.Theor.Phys.801}, and first-~\cite{Arima1954Prog.Theor.Phys.509} and second-order~\cite{Ichimura1965Nucl.Phys.401} configurations mixing corrections are important in improving the relativistic description of magnetic moments~\cite{Li2013Phys.Rev.C64307},  although such investigations so far have been mostly restricted to nuclei with doubly closed shells plus or minus one nucleon.
There are, of course, still many open questions, for instance, a) the role of the Dirac sea in the configuration mixing; b) the influence of higher order diagrams~\cite{Bauer1973Nucl.Phys.A535}; c) the coupling to the isobar current~\cite{Rho1974Nucl.Phys.A493}; d) the influence of the self-consistent inclusion of the pion and the tensor forces in the relativistic Hartree-Fock theory~\cite{Long2006Phys.Lett.B150}, etc. Work in these directions should be done in future.

On the other hand, the agreement between the PQTAC results and data may also partially from the inclusion of pairing correlations. It was found that pairing correlations affect the dynamics of the valence particles and/or holes near the Fermi surface and, thus, could reduce the $B(M1)$ values~\cite{Zhao2015Phys.Rev.C34319}. Finally, it should be noted that the shear bands are characterized by the decreasing tendency of the $B(M1)$ values, but not their absolute magnitudes.

\begin{figure}[ht]
\centering
\vspace{0mm}
\includegraphics[width=0.8\columnwidth]{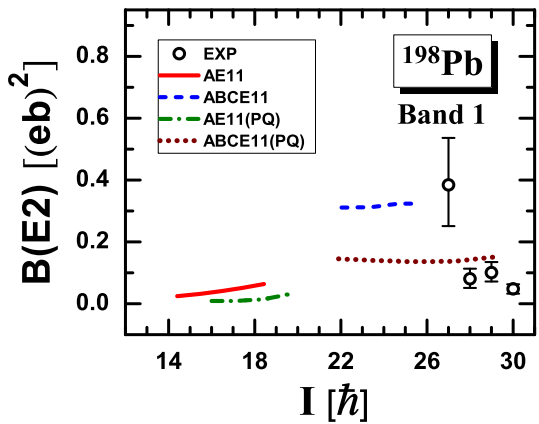}
\vspace{-4mm}
\caption{$B(E2)$ values as  functions of the total angular momentum in the TAC-CDFT calculations compared
with the data from Ref.~\cite{Clark1997Phys.Rev.Lett.1868} and the PQTAC results~\cite{Chmel2007Phys.Rev.C75} for band 1. Taken from Ref.~\cite{Yu2012Phys.Rev.C24318}.\label{fig3-7}}
\end{figure}

In magnetic rotations, the $E2$ transitions are weak in contrast to the strong $M1$ transitions.
In Fig.~\ref{fig3-7}, it is shown the comparison among the $B(E2)$ values from TAC-CDFT~\cite{Yu2012Phys.Rev.C24318} and PQTAC calculations~\cite{Chmel2007Phys.Rev.C75} as well as the corresponding data~\cite{Clark1997Phys.Rev.Lett.1868}.
It shows a reasonable agreement between the TAC-CDFT results and the data. Moreover, the TAC-CDFT results have a roughly constant tendency with the increasing spin, revealing almost constant deformation.
The TAC-CDFT calculations provide larger $B(E2)$ values than the PQTAC results. This is because, as illustrated in Ref.~\cite{Yu2012Phys.Rev.C24318}, larger deformations are obtain in the TAC-CDFT calculations.

It should be mentioned that the transition probabilities for shears bands in Pb isotopes from $^{193}$Pb to $^{202}$Pb were analyzed in detail with the PQTAC approach in Ref.~\cite{Chmel2007Phys.Rev.C75}.
The authors also compared their results with those from a geometrical
analysis~\cite{Clark2000Annu.Rev.Nucl.Part.Sci.1} with the shears angle, i.e., the angle between the proton and neutron angular momenta.
For the $B(M1)$ values, results from both approaches are in qualitative agreement with each other, and reproduce the observed trend of the data.
This provides a confirmation of shears mechanism in these bands. However, differences exist for the $B(E2)$ values. The PQTAC calculations predict a weak spin dependence for the $B(E2)$ values, while the geometrical approach provides a rapidly decreasing tendency. Although more accurate experimental data were needed to decide between the models, it was pointed out that the core polarization effects, which is not properly treated in the geometrical approach, play an important role in describing the $E2$ transitions. By fully taking into account the core polarization effects, the calculated $B(E2)$ values with CDFT show a similar weak spin dependence to the PQTAC results.

Apart from the magnetic rotation bands in Pb nuclei, the TAC-CDFT has also been applied successfully in describing the magnetic dipole bands in many others regions so far. Among these are the $A\sim60$ region~\cite{Zhao2011Phys.Lett.B181,Steppenbeck2012Phys.Rev.C44316}, the $A\sim80$ region~\cite{Madokoro2000Phys.Rev.C61301}, the $A\sim110$ region~\cite{Peng2015Phys.Rev.C44329}, the $A\sim140$ region~\cite{Peng2008Phys.Rev.C24313}, etc. These facts demonstrate that the TAC-CDFT has become a powerful theoretical tool for the nuclear magnetic rotations.

%%%%%%%%%%%%%%%%%%%%%%%%%%%%%%%%%%%%%%%%%%%%%%%%%%%%%%%%%%%%%%%%%%%%%%%%%%%%%%%%

\subsection{Antimagnetic Rotation}
\label{Sect.03.02}

As mentioned in the introduction, antimagnetic rotation occurs in nearly spherical nuclei.
A pair of valence particles or holes in high-$j$ orbitals have their angular momenta oppositely aligned at the bandhead.
The rotational symmetry is violated by the valence nucleons in such a system and antimagnetic rotation bands could be built~\cite{Frauendorf1996272,Frauendorf2001Rev.Mod.Phys.463}.
The first TAC-CDFT investigation for antimagnetic rotation was performed for the nucleus
$^{105}\rm Cd$~\cite{Zhao2011Phys.Rev.Lett.122501}.
In the following, we take this as an example to demonstrate the description of the AMR characteristics.
The calculations were performed with the functional PC-PK1~\cite{Zhao2010Phys.Rev.C54319}, and more details can be found in Refs.~\cite{Zhao2011Phys.Rev.Lett.122501,Zhao2012Phys.Rev.C54310}.

\subsubsection{Routhians}

\begin{figure}[ht]
\centering
\vspace{0mm}
\includegraphics[width=0.8\columnwidth]{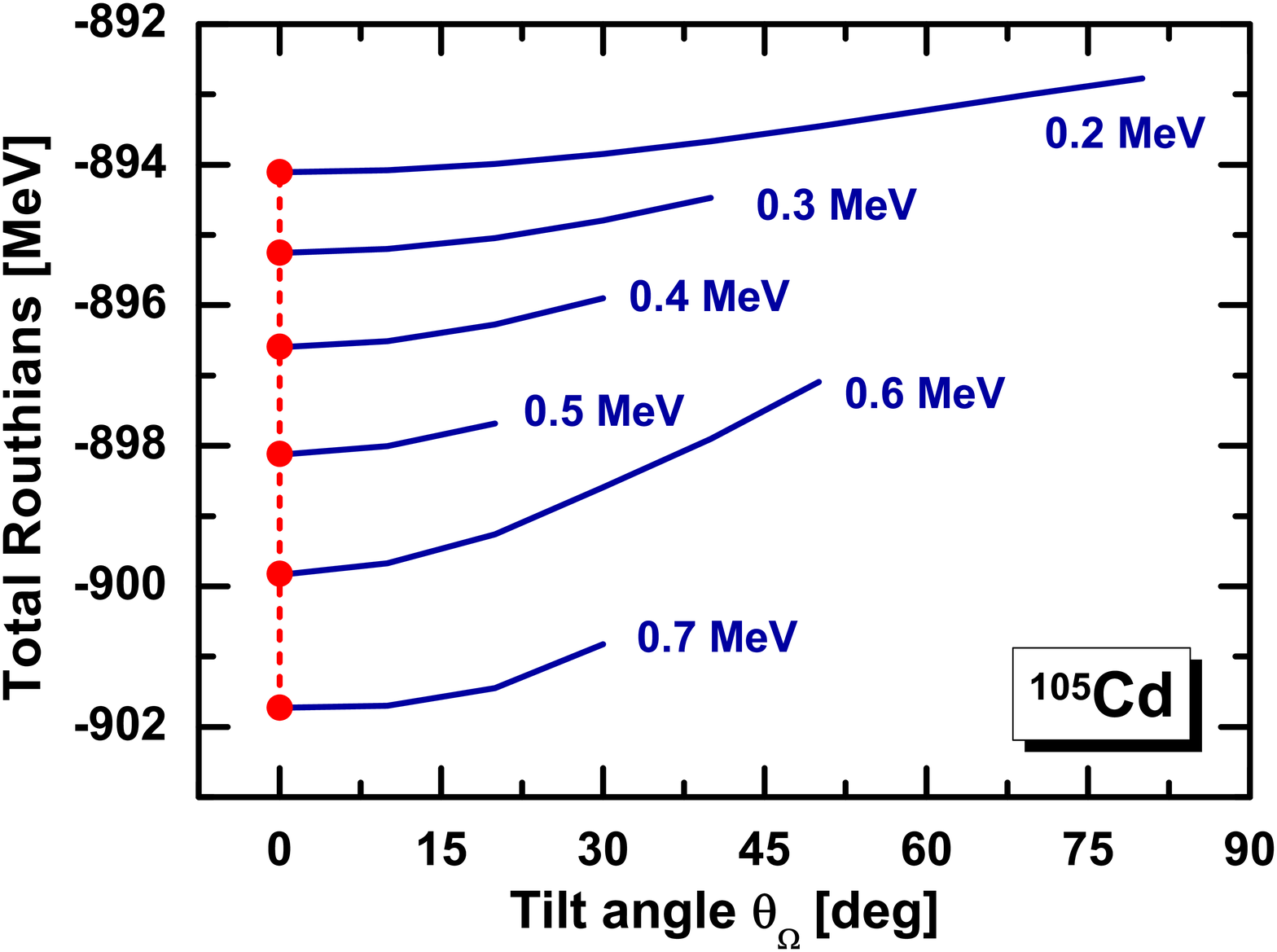}
\vspace{-4mm}
\caption{TAC-CDFT results for total Routhians (solid lines) as functions of the tilt angle $\theta_\Omega$. The solid dots represent the minima for the corresponding frequency. Taken from Ref.~\cite{Zhao2012Phys.Rev.C54310}.\label{fig4-1}}
\end{figure}

There are 48 protons and 57 neutrons in the nucleus $^{105}\rm Cd$. To describe the observed AMR band in this nucleus, one neutron particle is fixed in the lowest level of $h_{11/2}$ shell, while the remain nucleons are treated self-consistently by occupying the single-particle levels according to their energies.
This automatically leads to two proton holes in the $g_{9/2}$ shell, which are aligned equally in two opposite directions. For the neutron particles, apart from the one in the $h_{11/2}$ shell, there are six ones distributed over the $(g_{7/2}d_{5/2})$ shell with strong mixing with each other.

An important feature of an AMR rotor is that the system is symmetric under a rotation around the rotation axis by an angle of $\pi$. This means that the rotational axis must not be tilted, but parallel with one of the principal axes. In principle, the tilt angle $\theta_\Omega$, the angle between the rotational axis and the principal $x$ axis, should be determined in a self-consistent way by minimizing the total Routhian
\begin{equation}
  E'(\Omega,\theta_\Omega) = E_{\rm tot}-\langle \cos\theta_\Omega\Omega J_x+\sin\theta_\Omega\Omega J_z \rangle
\end{equation}
with respect to the angle $\theta_\Omega$.
In Fig.~\ref{fig4-1}, the total Routhians are shown as functions of the tilted angle $\theta_\Omega$.
It is very clear that the total Routhians for various rotational frequencies always have their mimima at $\theta_\Omega = 0^\circ$, which means that the rotational axis of antimagnetic rotation is the $x$ axis. This results from the symmetry for an antimagnetic configuration, but one should be noted that it is not trivial to obtain such results in the TAC-CDFT calculations, because in principle symmetries can be broken in the self-consistent solution.
Principal axis cranking calculations can also be used to describe the antimagnetic rotation band in nuclei, while TAC calculations can easily provide the relative angle of the $g_{9/2}$ proton hole vectors (see Fig.~\ref{fig4-3} below).

\subsubsection{Energy spectrum}

\begin{figure}[ht]
\centering
\vspace{0mm}
\includegraphics[width=0.8\columnwidth]{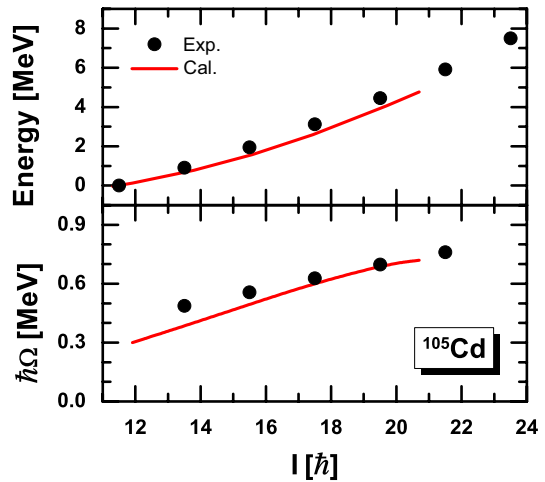}
\vspace{-4mm}
\caption{Rotational energy calculated (upper panel) and the rotational frequency (lower panel) versus angular momentum of the AMR band in $^{105}$Cd in comparison with data~\cite{Choudhury2010Phys.Rev.C61308} (solid dots). Taken from Ref.~\cite{Zhao2011Phys.Rev.Lett.122501}.\label{fig4-2}}
\end{figure}

In Fig.~\ref{fig4-2}, the rotational energy calculated and the rotational frequency (solid lines) versus angular momentum~\cite{Zhao2011Phys.Rev.Lett.122501} are compared with data~\cite{Choudhury2010Phys.Rev.C61308}. The experimental rotational excitation energies are in an excellent agreement with the self-consistent TAC-CDFT calculations. In the lower panel, the total angular momenta are also reproduced quite well with the data. In particular, it should emphasize that the angular momenta are increased almost linearly with rotational frequency in this AMR band, i.e., the moment of inertia is almost a constant. The fact that TAC-CDFT calculations provide a very good description of the angular momentum reveals that the nearly constant moment of inertial could also be well reproduced.

\subsubsection{Two shears-like mechanism}

\begin{figure}[ht]
\centering
\vspace{0mm}
\includegraphics[width=0.8\columnwidth]{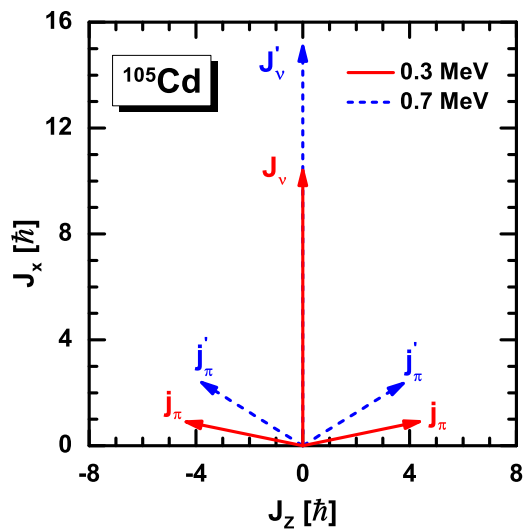}
\vspace{-4mm}
\caption{Angular momentum vectors of neutrons $\bm{J}_\nu$ and the two $g_{9/2}$ proton holes $\bm{j}_\pi$ at both the bandhead ($\hbar\Omega=0.3$~MeV)
and the maximum rotational frequency.  Taken from Ref.~\cite{Zhao2011Phys.Rev.Lett.122501}.\label{fig4-3}}
\end{figure}

The two shears-like mechanism are illustrated in Fig.~\ref{fig4-3} with displaying the angular momenta of the two $g_{9/2}$ proton holes $\bm{j}_\pi$ and neutron angular momentum $\bm{J}_\nu$ at both the bandhead and the maximum rotational frequency.
At the bandhead, the two proton angular momenta $\bm{j}_\pi$ align in two opposite directions, and they are nearly perpendicular to the neutron angular momentum $\bm{J}_\nu$. With increasing rotational frequency, the angular momentum is increased by simultaneously aligning the angular momenta of the two $g_{9/2}$ proton holes toward the neutron one $\bm{J}_\nu$. The direction of the total angular momentum keeps always along the $x$ axis. This forms an analogy to the closing of the blades of two shears and, thus, the two shears-like mechanism is clearly seen.

\begin{figure}[ht]
\centering
\vspace{0mm}
\includegraphics[width=0.8\columnwidth]{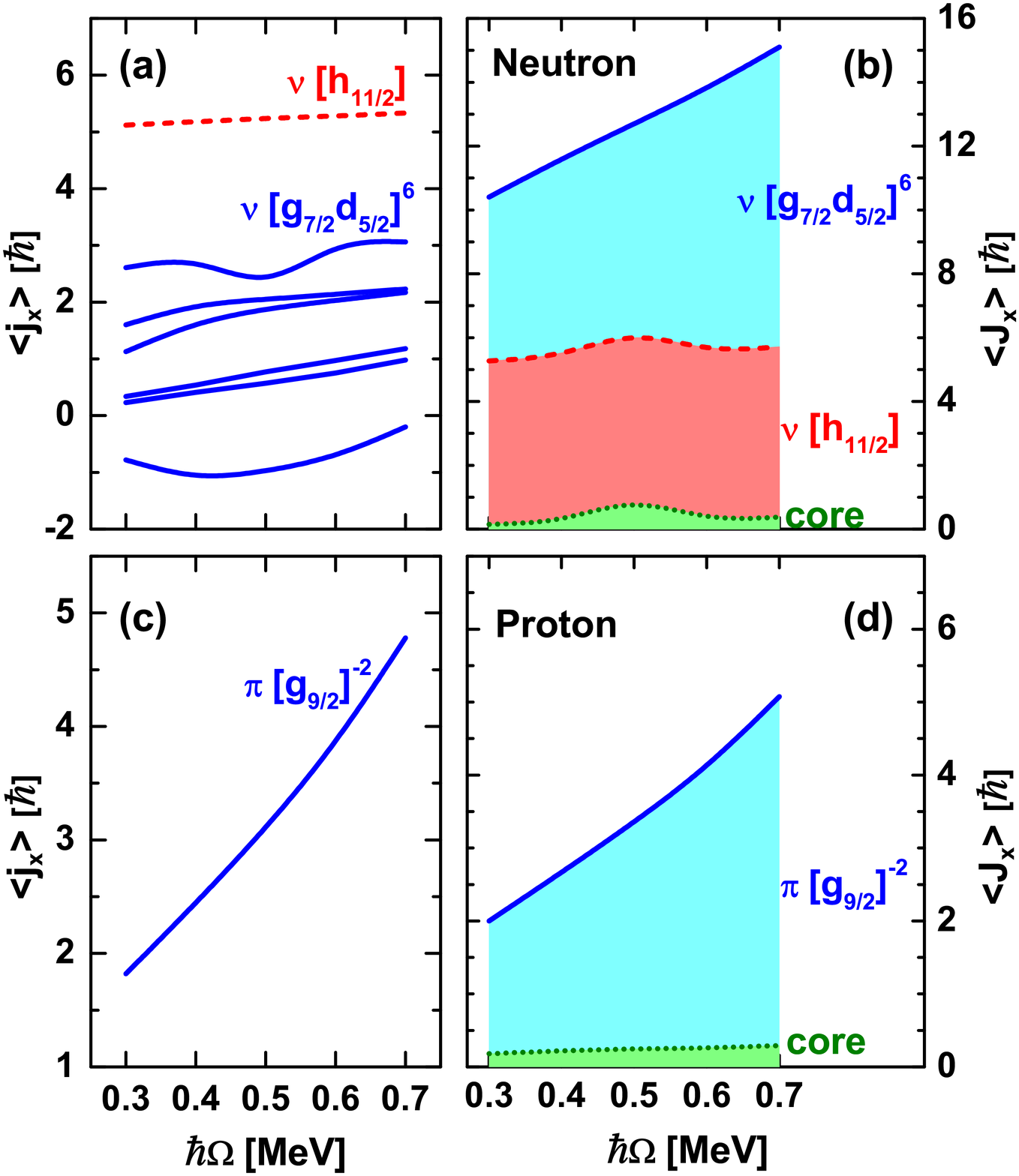}
\vspace{-4mm}
\caption{Angular momentum alignment for the seven neutron particles [panel (a)]  and two proton holes [panel (c)] as well as the compositions of the neutron [panel (b)] and proton [panel (d)] angular momenta. Taken from Ref.~\cite{Zhao2012Phys.Rev.C54310}.\label{fig4-4}}
\end{figure}

In a fully self-consistent and microscopic calculation, there is no inert core. All the energy and angular momentum come from individual nucleons.
The microscopic picture of the two shearlike mechanism are shown in Fig.~\ref{fig4-4}.
For the nucleus $^{105}$Cd, one can easily define a $^{100}$Sn core with 50 protons and 50 neutrons occupied below a close shell. The angular momentum from this such a core is quite small.
Instead, the angular momentum is mainly generated from the valence nucleons in the high-$j$ orbitals including the two proton holes in the $g_{9/2}$ shell and the seven neutron particles in the $h_{11/2}$ and $g_{7/2}$ shells.

For the protons, one can see the angular momentum alignment and the compositions of the total proton angular momentum in the panel (c) and panel (d), respectively.
The proton angular momentum is mainly from the two holes in $g_{9/2}$ shell. The angular momentum of two proton holes cancel with each other in the $z$ direction.
With the increasing rotational frequency, the proton angular momentum is increased in the $x$ direction by aligning the two proton holes towards the rotating axis. This corresponds to the two shears-like mechanism.

For the neutrons, one can see the angular momentum alignment and the compositions of the total neutron angular momentum in the panel (a) and panel (b), respectively.
The neutron angular momentum is mainly from the neutron particles in the $h_{11/2}$ and $g_{7/2}$ shell.
There is always a neutron in the $h_{11/2}$ orbital, which contributes to the neutron angular momentum of roughly $5\hbar$. This amount of $5\hbar$ barely change when the rotational frequency increases. Instead, the neutron angular momentum is mainly increased by aligning other six neutrons in the $g_{7/2}$ and $d_{5/2}$ shells.

This microscopic calculation shows that there are clearly two proton holes and one neutron particle in the $g_{9/2}$ and the $h_{11/2}$ orbitals, respectively.
However, other valence neutrons are distributed over the lower-$j$ orbitals above the $N=50$ core because of the strong mixing between different orbitals.
Along the band, the angular momentum generates by aligning the two proton holes and also the mixing within the neutron orbitals.
Due to this strong mixing, a core frozen in the phenomenological model in Ref.~\cite{Clark2000Annu.Rev.Nucl.Part.Sci.1} is not appropriate.

It should be noted that here the mixing indeed results from the spontaneous symmetry breaking. In the TAC-CDFT calculations, the particles are moving in a rotating deformed potential on given single-particle orbitals.
Such orbitals are usually distributed over spherical orbitals with the quantum numbers ($ n\ell j j_x$). Here, $n$ is the radial quantum number, ($\ell j$) are the quantum numbers of angular momentum and $j_x$ is the magnetic quantum number in the $x$ direction. With the increasing rotational frequency, the alignment of the angular momentum is increased by mixing  single-particle orbitals with different quantum numbers ($ n\ell j j_x$).

\subsubsection{Electric transition and deformation}

\begin{figure}[ht]
\centering
\vspace{0mm}
\includegraphics[width=0.8\columnwidth]{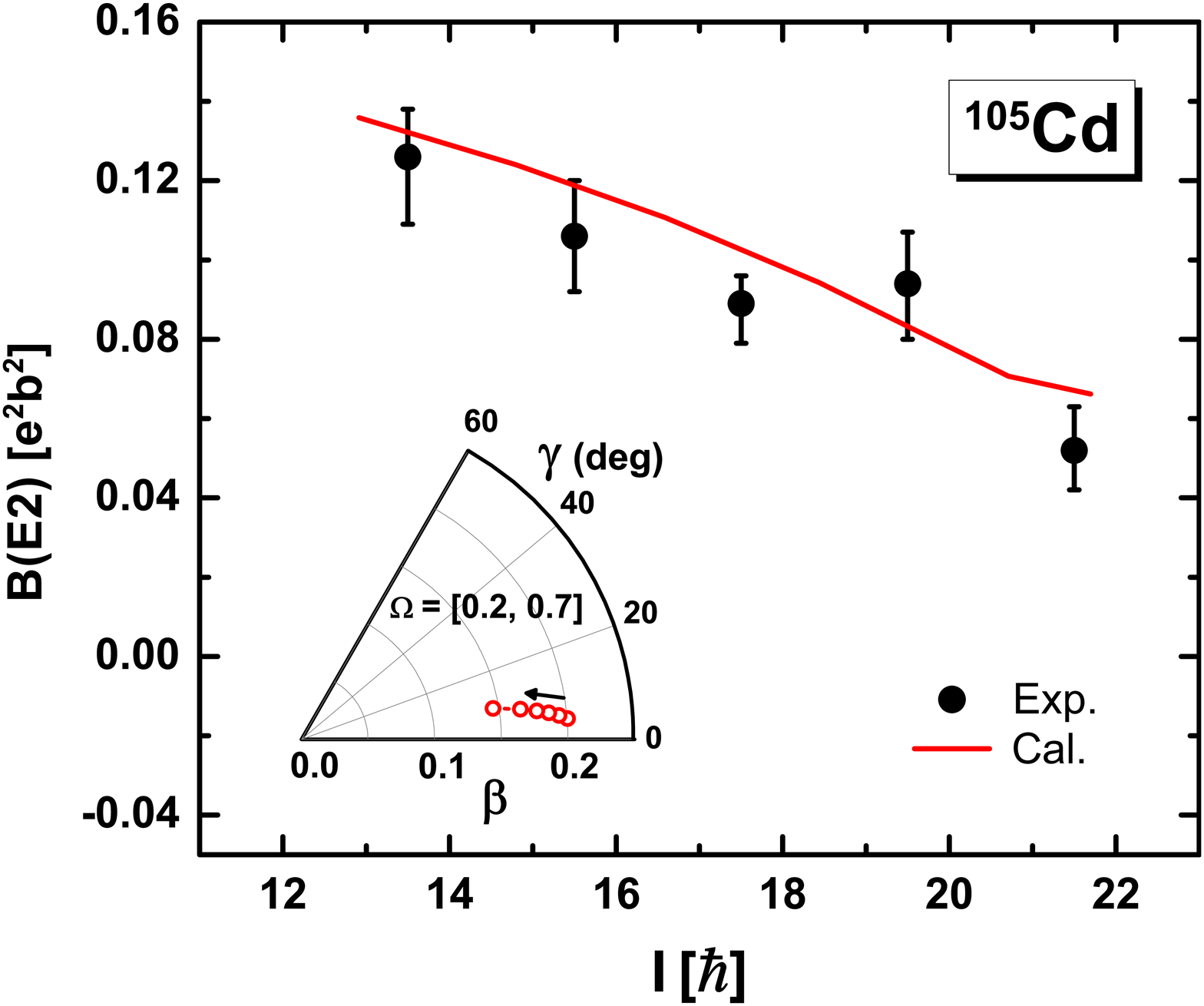}
\vspace{-4mm}
\caption{Calculated $B(E2)$ values in comparison with data~\cite{Choudhury2010Phys.Rev.C61308} (solid dots). Inset: Deformations $\beta$ and $\gamma$ driven by the increasing rotational frequency whose direction is indicated by arrows.  Taken from Ref.~\cite{Zhao2011Phys.Rev.Lett.122501}.\label{fig4-5}}
\end{figure}

Antimagnetic rotation is characterized by the decreasing tendency of $E2$ transitions. In Fig.~\ref{fig4-5}, the calculated $B(E2)$ values from TAC-CDFT are compared with the corresponding data~\cite{Choudhury2010Phys.Rev.C61308}. The $B(E2)$ values are very small ($< 0.14~e^2b^2$) and are reproduced very well by the calculations.
Moreover, it also shows decreasing tendency of the $B(E2)$ values, which is consistent with the two shears-like mechanism.

This decreasing tendency can be understood with the deformation evolution.
As shown in the inset of Fig.~\ref{fig4-5}, the $\beta$ deformation decreases rapidly with increasing frequency, while the $\gamma$ deformation is small and nearly unchanged.
The $B(E2)$ values are closely related to the deformation of the charge distribution.
As addressed in Ref.~\cite{Zhao2012Phys.Rev.C54310}, the deformation of the charge distribution changes in a similar manner to the total density deformation.
Therefore, one has the conclusion that the two shears-like mechanism, here the simultaneous alignment of the two proton holes, results in a transition from prolate shape to nearly spherical shape.

\subsection{Chiral and Multi-Chiral Rotation}
\label{Sect.03.03}

\begin{figure*}[ht]
\centering
\vspace{0mm}
\includegraphics[width=0.8\columnwidth]{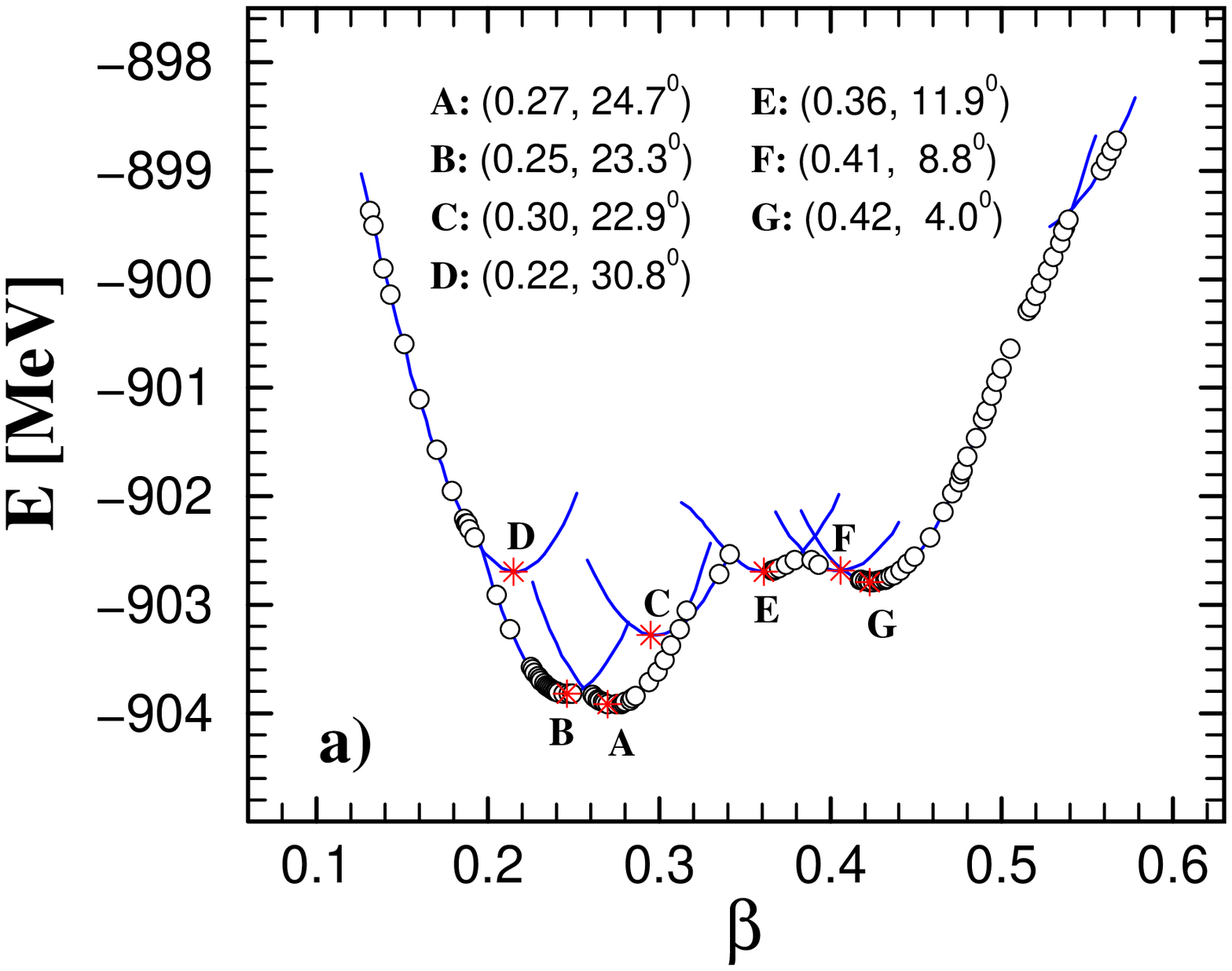}
\includegraphics[width=0.8\columnwidth]{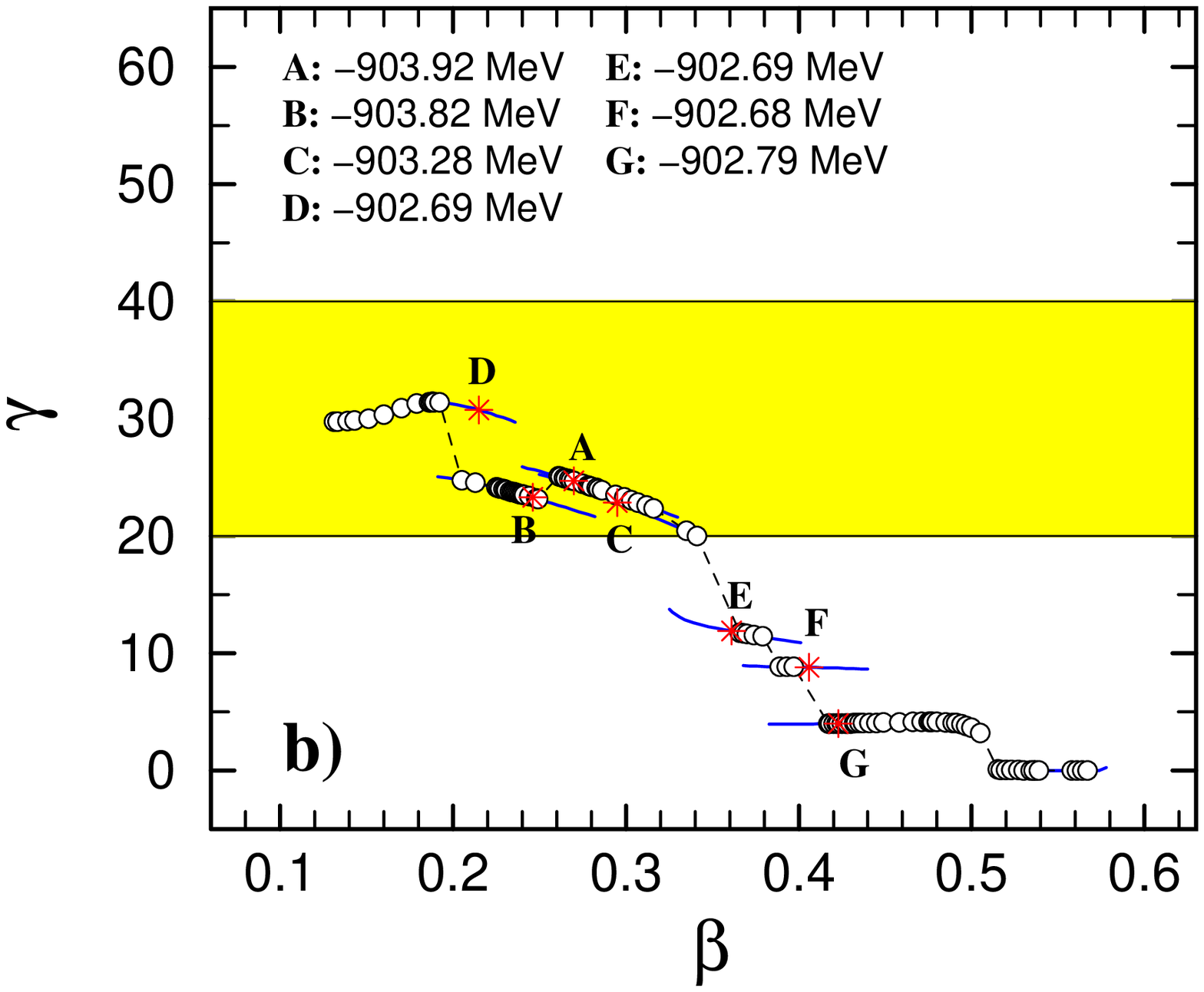}
\vspace{-4mm}
\caption{ The energy surfaces (a) and the $\gamma$ deformations (b) as functions of $\beta$ deformation in adiabatic (open circles) and configuration-fixed (solid lines) constrained triaxial CDFT calculation for $^{106}$Rh. The minima in the energy surfaces are represented as stars and labeled respectively as A, B, C, D, E, F and G. Their corresponding deformations $\beta$ and $\gamma$ together with their energies are respectively given in (a) and (b). Taken from Ref.~\cite{Meng2006Phys.Rev.C37303}.\label{fig5-1}}
\end{figure*}

Frauendorf and Meng originally suggested the existence of spontaneous chiral symmetry breaking in rotating triaxial odd-odd nuclei in 1997~\cite{Frauendorf1997Nucl.Phys.A131}. Since then, the chiral doublets bands have been investigated by many approaches including the triaxial particle rotor model (PRM)~\cite{Frauendorf1997Nucl.Phys.A131,Peng2003Phys.Rev.C44324,Zhang2007Phys.Rev.C44307,Qi2009Phys.Lett.B175}, the interacting boson fermion-fermion model
(IBFFM)~\cite{Tonev2006Phys.Rev.Lett.52501,Tonev2007Phys.Rev.C44313,Brant2008Phys.Rev.C34301}, the tilted axis cranking mean field with shell correction model (SCTAC)~\cite{Dimitrov2000Phys.Rev.Lett.5732}, tilted axis cranking Skyrme-Hartree-Fock model~\cite{Olbratowski2004Phys.Rev.Lett.52501,Olbratowski2006Phys.Rev.C54308}, and the random phase approximation~\cite{Mukhopadhyay2007Phys.Rev.Lett.172501,Almehed2011Phys.Rev.C54308}.
Very recently, the chiral rotation and vibration are also investigated with a new collective Hamiltonian~\cite{Chen2013Phys.Rev.C24314}. Similar model has also been extended to describe the wobbling mode in triaxial nuclei as well~\cite{Chen2014Phys.Rev.C44306}.

Based on the covariant density functional theory, one can introduce the three-dimensional cranking approach to investigate the chiral rotation self-consistently. Due to the numerical complexity, however,  such calculations has not been carried out yet.
Nevertheless, it is constructive to apply the self-consistent CDFT to search for nuclear configurations and triaxial deformation parameters suitable for chirality.
For this purpose, the adiabatic and configuration-fixed
constrained triaxial CDFT approaches were first used to investigate the triaxial shape
coexistence and possible chiral doublet bands in $^{106}$Rh~\cite{Meng2006Phys.Rev.C37303}.
It was demonstrated that multiple pairs of chiral doublet bands could exist in a single
nucleus, which is the so-called M$\chi$D~\cite{Meng2006Phys.Rev.C37303}.
A series of works following this direction has been carried out by either examining the influence of other effects like the time-odd effects~\cite{Yao2009Phys.Rev.C67302}, or searching for other possible candidates like
$^{104,106,108,110}$Rh~\cite{Peng2008Phys.Rev.C24309}, $^{105}$Rh~\cite{Li2011Phys.Rev.C37301}, and $^{107}$Ag~\cite{Qi2013Phys.Rev.C27302}. The likelihood of chiral bands with different configurations was also discussed in the TAC calculations for the bands observed in $^{105}$Rh~\cite{Timar2004Phys.Lett.B178,Alcantara-Nunez2004Phys.Rev.C24317}. The first experimental evidence for the M$\chi$D was reported in 2013 in $^{133}$Ce~\cite{Ayangeakaa2013Phys.Rev.Lett.172504}.

Moreover, the robustness of chiral geometry against the increase
of the intrinsic excitation energy has been investigated theoretically in Refs.~\cite{Droste2009Eur.Phys.J.A79,Chen2010Phys.Rev.C67302,Hamamoto2013Phys.Rev.C24327}. It turns out that the chiral geometry could be sustained in the higher excited bands of a certain chiral configuration, i.e., M$\chi$D based on the same configuration. Very recently, the first experimental evidence
for such a new type of M$\chi$D has been reported in $^{103}$Rh~\cite{Kuti2014Phys.Rev.Lett.32501}.

\subsubsection{Multiple chiral bands (M$\chi$D)}

Triaxial CDFT calculations was firstly carried out for $^{106}$Rh in Ref.~\cite{Meng2006Phys.Rev.C37303} to search for the possible M$\chi$D, which is closely related to the triaxial shape coexistence in nuclei.
To check the triaxial shape coexistence in the CDFT, in principle, one needs to obtain the energy surface with respect to the deformation by the two-dimensional ($\beta$ and $\gamma$) constrained calculations~\cite{Ring1980}. However, this could be very time-consuming and, thus, for simplicity an alternative $\beta^2$ constrained calculations were carried out to search for the triaxial states in nuclei in Ref.~\cite{Meng2006Phys.Rev.C37303}.

The obtained energy surface and the deformation $\gamma$ for $^{106}$Rh are presented as functions of
deformation $\beta$ in Fig.~\ref{fig5-1}(a) and (b), respectively. The open circles show the results from the adiabatic constrained calculations.
Since the configurations could be changed in these results, it exhibits some irregularities in the energy surface, and some energy minima are even too obscure to be recognized. To overcome these shortcomings, the configuration-fixed calculation~\cite{Simon1983Phys.Rev.Lett.2167,Bengtsson1989Nucl.Phys.A56,Guo2004Nucl.Phys.A59} was introduced in the triaxial CDFT calculations~\cite{Meng2006Phys.Rev.C37303}. The corresponding results are shown in Fig.~\ref{fig5-1}(a) and (b) as solid lines.
For each fixed configuration, the constrained calculation gives a
continuous and smooth curve for the energy surface and the $\gamma$ deformation.
Now the energy minima could be identified clearly, and they are represented by stars and labeled as A, B, C, D, E, F and G.

For each fixed configuration, the $\gamma$ deformation is approximately constant [see Fig~\ref{fig5-1}(b)],
indicating that the triaxial deformation is relatively stable. The nucleus $^{106}$Rh
provides a good example of the triaxial shape coexistence; the energies differences are within 1.3 MeV but they correspond to quite different $\beta$ and $\gamma$ deformations. In addition, the configurations A, B, C, and D have considerable $\gamma$ deformation suitable for chirality. In addition to the $\gamma$ deformation, one must have high-$j$ particles and holes to build the chiral geometry.
This was found for the ground state (state A) as $\pi(1g_{9/2})^{-3} \otimes \nu (1h_{11/2})^2 $,
state B as $\pi(1g_{9/2})^{-3} \otimes \nu (1h_{11/2})^1 $, and state C as $\pi(1g_{9/2})^{-3} \otimes \nu(1h_{11/2})^3$, respectively. For the state D, however, there is no high-$j$ neutron valence
particle. The configurations A, B, and C possess high-$j$ proton holes, high-$j$
neutron particles, and considerable $\gamma$ deformations, and thus favor the construction of the chiral doublet bands. This leads to the prediction of the M$\chi$D in Ref.~\cite{Meng2006Phys.Rev.C37303}.

\subsubsection{M$\chi$D with distinct configurations: $^{133}$Ce}

Since the prediction of M$\chi$D, lots of experimental effort
have been made to search for the corresponding evidences. Using the ATLAS facility
at the Argonne National Laboratory, high-spin states in $^{133}$Ce were populated following
the $^{116}$Cd($^{22}$Ne, 5$n$) reaction. Among the observed rotation bands, two sets of doublet bands (labeled as bands 5 and 6 and bands 2 and 3 respectively) were distinctly identified~\cite{Ayangeakaa2013Phys.Rev.Lett.172504}.

\begin{figure}[ht]
\centering
\vspace{0mm}
\includegraphics[width=0.95\columnwidth]{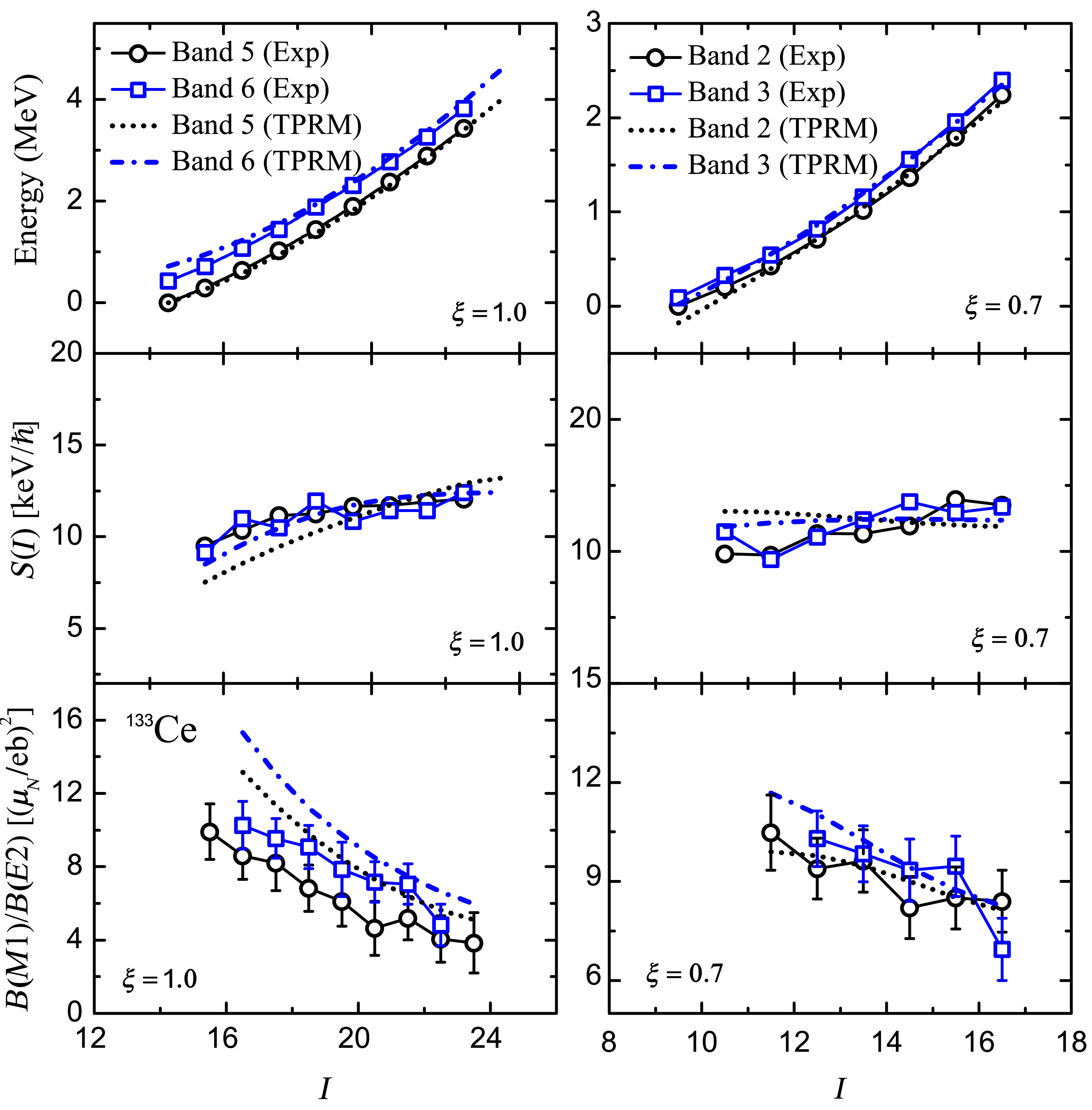}
\vspace{-4mm}
\caption{Experimental excitation energies, $S(I)$ parameters, and $B(M1)/B(E2)$ ratios for the negative-parity chiral doublet (left panels) and positive-parity chiral doublet (right panels) in $^{133}$Ce. Also shown are results of PRM calculations with the indicated attenuation factors $\xi$ (see text). Taken from Ref.~\cite{Ayangeakaa2013Phys.Rev.Lett.172504}.\label{fig5-2}}
\end{figure}

Despite the difficulties in the self-consistent three-dimensional TAC-CDFT calculations, one can combine the constrained triaxial CDFT and the triaxial particle rotor model to investigate the observed pairs of doublet bands in $^{133}$Ce. In such calculations, the obtained deformations and configurations from CDFT are regarded as input for the triaxial particle rotor model, and then the results obtained from the latter will be compared with the data.
In Fig.~\ref{fig5-2}, the calculated energy spectra, $S(I)$ parameters, and $B(M1)/B(E2)$ ratios are compared
with the experimental values for the negative parity bands 5 and 6 as well as
the positive parity bands 2 and 3.
The configuration of $\pi(1h_{11/2})^2\otimes \nu(1h_{11/2})^{-1}$ is assigned for bands 5 and 6.
The energy differences between these two bands are nearly constant of around 400 keV, and the calculated results could reproduce the data quite well.
It was suggested a form of chiral vibration for these two bands, i.e., a tunneling between the left- and right-handed
configurations. In such case, bands 5 and 6 correspond to the zero- and one-phonon states, respectively.
Moreover, the calculated staggering parameter $S(I)$ varies smoothly with the angular momentum.
This is because that the Coriolis effect
is considerably reduced for a three-dimensional coupling of angular momenta in a chiral geometry.
No apparent odd-even $B(M1)/B(E2)$ staggering are shown in the calculated results, although a small effect is exhibited in the experimental data.
This further supports the chiral vibration interpretation between bands 5 and 6~\cite{Qi2009Phys.Lett.B175}.

The bands 2 and 3 with positive parity are built based on the configuration
$\pi[(1g_{7/2})^{-1}(1h_{11/2})^1]\otimes \nu(1h_{11/2})^{-1}$.
In the calculations, since the low-$j$ orbital $\pi(1g_{7/2})$ in the configuration could have large mixing with other low-$j$ orbitals, a Coriolis attenuation factor $0.7$ was employed to describe the data.
The energy differences between the corresponding states in bands 2 and 3 are nearly
constant around $100$ keV. Moreover, the staggering parameter $S(I)$ barely change with the angular momentum.
All of these are consistent with the chiral interpretation of these bands. For bands 5 and 6,
similarly, there is no apparent staggering in the calculated $B(M1)/B(E2)$ ratios, while there is staggering in the experimental ones.
As the $B(M1)/B(E2)$ ratio sensitively depends on the details of the transition
from the vibrational to the tunneling regime~\cite{Qi2009Phys.Lett.B175}, this may partly
account for the deviations between the calculated and experimental ones.

\subsection{M$\chi$D with identical configuration: $^{103}$Rh}
Frauendorf and Meng studied the system of a high-$j$ particle and a high-$j$ hole coupled to a triaxial rotor in Ref.~\cite{Frauendorf1997Nucl.Phys.A131}.
As shown in Fig.~\ref{fig1-4}, the near degeneracy occurs not only for the two lowest bands, but also for higher excited bands. This indicates that chiral geometry may be remained in these excited bands as well.
Sequentially, it has also been demonstrated in other theoretical calculations~\cite{Droste2009Eur.Phys.J.A79,Chen2010Phys.Rev.C67302,Hamamoto2013Phys.Rev.C24327} that the phenomenon of M$\chi$D may exist with same particle-hole configuration, i.e., not only between the yrast and yrare bands but also between two higher excited bands. This has been observed in the recent experiment of $^{103}$Rh, where three bands with positive parity and five ones with negative parity were constructed~\cite{Kuti2014Phys.Rev.Lett.32501}.

To understand the observed band structure in $^{103}$Rh, one needs first to explore the possible configurations.
This were carried out with the adiabatic and configuration-fixed constrained CDFT calculations~\cite{Meng2006Phys.Rev.C37303}.
Subsequently, the obtained configurations and the corresponding deformations were further confirmed and checked with the two-dimensional TAC-CDFT calculations~\cite{Zhao2011Phys.Lett.B181}.
The energy spectra, Routhians, angular momentum, deformations, and single-particle alignments were analyzed.
Finally, the confirmed configurations and deformations were adopted as the inputs of
quantum triaxial particle rotor model~\cite{Zhang2007Phys.Rev.C44307}, and the obtained results were used to
explain both the positive- and negative-parity bands.

\begin{figure*}[ht]
\centering
\vspace{0mm}
\includegraphics[width=1.4\columnwidth]{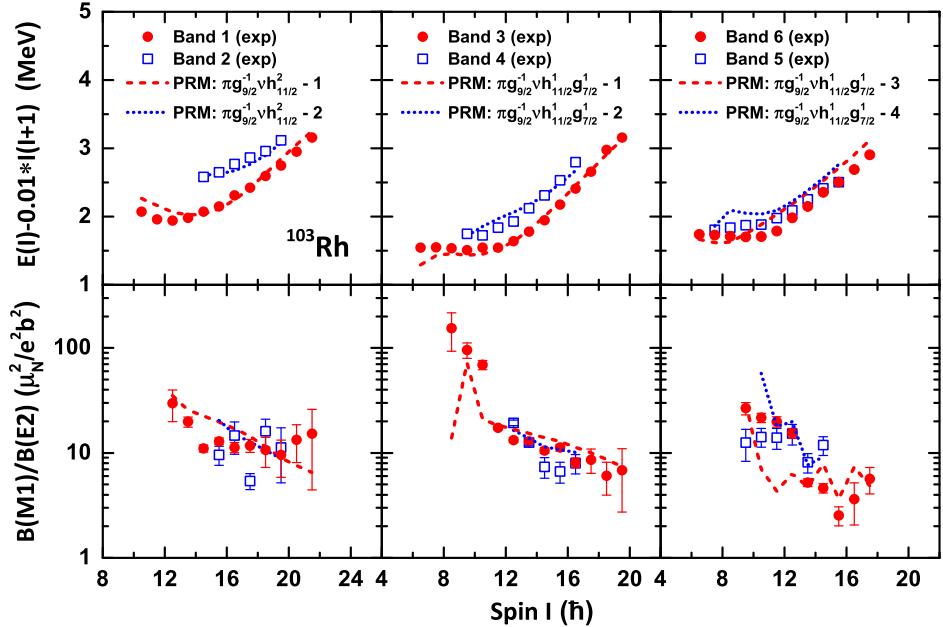}
\vspace{-4mm}
\caption{Experimental excitation energies and $B(M1)/B(E2)$ ratios for the positive-parity chiral bands 1-2 (left panels) and negative-parity multiple chiral bands 3-6 (middle and right panels) in $^{103}$Rh together with the results of triaxial particle rotor model. The number following the configuration label of the theoretical curve corresponds to the energy ordering of the calculated band with the given configuration. Taken from Ref.~\cite{Kuti2014Phys.Rev.Lett.32501}. \label{fig5-3}}
\end{figure*}

The calculated energy spectra are compared with the corresponding data in Fig.~\ref{fig5-3}.
For bands 1 and 2, the calculated results are in excellent agreement with the data. The energy separation of these two bands are roughly $\sim$500 keV at $I =29/2~\hbar$, and this value decreases with increasing angular momentum, finally drops to $\sim 360$ keV at $I=39/2$.
The $B(M 1)/B(E2)$ values of bands 1 and 2 are quite similar.
The calculations show apparent decreasing tendency for $B(M1)/B(E2)$ values, while such behavior is not observed in the experimental ratios. This may result from the frozen rotor assumption
adopted in particle rotor model.
As discussed in Ref.~\cite{Qi2011Phys.Rev.C34303}, the lower spin part of these two bands corresponds to chiral vibration, and with increasing spin, the static chirality comes out at the angular momentum
$I = 37/2$. Above this spin, the chiral mode tends to change back to another type of chiral vibration.

Coriolis attenuation was also employed to the four negative-parity bands 3-6 with the configuration
$\pi(1g_{9/2})^{-1}\otimes$ $\nu(1h_{11/2})^{1}$$(1g_{7/2})^{1}$~\cite{Kuti2014Phys.Rev.Lett.32501}.
The four calculated bands form two pairs of chiral doublets.
The first doublets well reproduce the experimental data of bands 3 and 4, and the second doublets
give reasonable agreement with the data of bands 6 and 5.
Quantitatively, the calculated energies for bands 5
and 6 are higher than the experimental values about 200 keV.
This is because that complex correlations are missing in the particle rotor model with
single-$j$ shell Hamiltonian.
The calculated electromagnetic transition probabilities are also in reasonable agreement with the data.
As discussed in Ref.~\cite{Qi2009Phys.Rev.C41302}, the weak odd-even $B(M1)/B(E2)$ staggering for bands 3 and 4 is consistent with the prediction of chiral vibration. For bands 5 and 6, the particle rotor model calculations reproduce also the staggering at $I=15.5~\hbar$.

In contrast with the M${\chi}$D in $^{133}$Ce, the observed negative-parity M${\chi}$D in $^{103}$Rh is built from the same configuration.
This observation shows that the chiral geometry in nuclei is robust against the intrinsic excitation.

%%%%%%%%%%%%%%%%%%%%%%%%%%%%%%%%%%%%%%%%%%%%%%%%%%%%%%%%%%%%%%%%%%%%%%%%%%%%%%%%
\section{Summary and perspective}
\label{Sect.04}

In the past decades, the rotational-like sequences in near-spherical or weakly deformed nuclei bring us to a new era of nuclear physics.
The phenomenon, known as magnetic rotation, has been extensively explored experimentally and theoretically. Subsequently, other modes, such as the antimagnetic rotation, and nuclear chirality and multi-chirality has been predicted theoretically and later confirmed in experiment.
The covariant density functional theory, with its many success in describing nuclear phenomena in both stable and exotic nuclei all over the nuclear chart, has been generalized to describe these rotational modes in nuclei.
In particular, the newly developed tilted axis cranking covariant density functional theory based on point-coupling interactions includes significant improvements which reduces the computation time and makes systematic investigation possible.

It should be mentioned that a proper treatment of time-odd fields is crucial for nuclear rotations. The nuclear density functionals are usually adjusted to experimental ground state properties or to nuclear matter data, which depend only on time-even parts of the potentials. Therefore, ambiguity for the time-odd part exists in non-relativistic density functionals. However, in relativistic functionals, the time-odd fields described by the currents, i.e., the spatial components of the vector fields, have the same coupling constant as the time-like components and don't require new parameters.

So far, the tilted axis cranking covariant density functional theory has achieved great success in explaining a number of magnetic rotational bands occurring in nuclei at the $A\sim60$, $A\sim80$, $A\sim110$, $A\sim140$, and $A\sim190$ mass regions. This indicates that a relativistic description for nuclear magnetic rotation is applicable for a large scope spread over the nuclear chart.
It turns out that such relativistic approach also provides a very good description for nuclear antimagnetic rotation, where its microscopic picture and interplay between collective rotation can be clearly demonstrated in a self-consistent way. The multi chiral doublet bands, more than one pair of chiral doublet bands in a single nucleus, has been first predicted based on the adiabatic and configuration-fixed
constrained relativistic density functional calculations. The experimental evidence for this prediction has been found in the recent data.
These facts show that the relativistic models have been very successful in describing nuclear titled axis rotations, and they are promising tools to bring us new exciting results in the future.

Despite the achievement, there are still many interesting topics and open problems in this field. In the following, we briefly discuss several of them.

{\it Pairing effects.} The pairing correlations have been neglected in most tilted axis cranking covariant density functional calculations so far. The pairing effects could be reduced by the blocked high $j$ oribitals near the Fermi surface in  magnetic and antimagnetic bands.
However, in many cases, the low $j$ orbitals also play an important role and, thus, the pairing correlations could be important.
In particular, for the magnetic rotation, recent calculations have found that the pairing would influence not only the magnitude of the angular momentum but also its direction~\cite{Zhao2015Phys.Rev.C34319}.
Due to the time-reversal symmetry broken in rotating nuclei, the conventional BCS method is
not valid and one should resort to the Bogoliubov method for the treating pairing. The most recent progress in this way can be found Ref.~\cite{Zhao2015Phys.Rev.C34319}.

Another way to treat the pairing correlations is to diagonalize a pairing Hamiltonian exactly a truncated shell model space. Since the particle number is exactly conserved in this method, it is thus known as Particle Number Conservation (PNC) method~\cite{Zeng1983Nucl.Phys.A1,Zeng1994Phys.Rev.C1388} in literature.
This method has also been implemented to treat the pairing channel in the framework of covariant density functional theory~\cite{MENG2006Front.Phys.China38}, known also as the Shell-model-like Approach (SLAP).
Similar calculations based on Nilsson levels were performed for antimagnetic rotation bands in Ref.~\cite{Zhang2013Phys.Rev.C54314}.
Therefore, it will be very interesting to implement SLAP to the present TAC covariant density functional theory for the pairing correlations.
In addition, the comparison between the SLAP results and those from the traditional Bogoliubov method for tilted axis rotation bands would also be very interesting.

{\it Aplanar cranking.}  To describe the novel chiral rotation in nuclei self-consistently, the three-dimensional cranking density functional theories are necessary. Such investigations have been
carried out with a mean-field combining a spherical Woods-Saxon potential with
a deformed Nilsson potential~\cite{Dimitrov2000Phys.Rev.Lett.5732} and the self-consistent mean field
from Skyrme-Hartree-Fock calculations~\cite{Olbratowski2004Phys.Rev.Lett.52501}.
For the relativistic case, no three-dimensional cranking calculations have been performed yet.
It would be interesting to combine the three-dimensional cranking approach with the point-coupling density functional, to simplify the numerical complicity and give self-consistent solutions for the chiral rotation.

{\it Beyond mean-field effects.}
For the present cranking relativistic models, the mean field approximation brings in the spontaneous symmetry breaking and, normally, the angular momentum is not good quantum number. This hinders an quantum description of the electromagnetic transition probabilities, e.g., $B(M1)$ and $B(E2)$. Moreover, a full description of chiral doublet bands also requires a theory going beyond the mean field approximation to restore the broken chiral symmetry in intrinsic frame.
Only aplanar cranking calculations can not describe the quantum tunneling between the left-handed and right-handed states in the body-fixed frame and would not yield the chiral partners.
The tilted axis cranking model with pairing plus quadrupole Hamiltonian has been extended by the random phase approximation~\cite{Mukhopadhyay2007Phys.Rev.Lett.172501}, but it limits to only the chiral vibration mode. New techniques tackling a unified description of chiral vibration and chiral rotation mode, at the moment, are still in their infancy~\cite{Chen2013Phys.Rev.C24314}.

{\it Other novel phenomena.}
In accompany with the worldwide development of the facilities of the radioactive ion beams, more and more exciting phenomena in nuclear states with spin degrees of freedoms would be observed. It will be definitely interesting to apply the relativistic approaches for understanding the existing data and anticipating the output of future experiments. For instance, a relativistic description for the recently observed nuclear ``tidal waves'', low-lying collective excitations may be described semiclassically as quadrupole running waves on the surface of the nucleus~\cite{PhysRevLett.110.102501}, will be quite helpful to deepen the understanding of this phenomenon. Moreover, novel deformation is always associated with novel rotation mode. In a recent study, the relativistic cranking calculations have been performed for carbon isotopes towards extreme spin and isospin~\cite{Zhao2015Phys.Rev.Lett.22501}. It was found that the coherent effects between the spin and isospin are helpful to stabilize the novel linear chain structure of three alpha clusters. Further investigations following this direction will be also important for the search of the corresponding experimental evidences.

%%%%%%%%%%%%%%%%%%%%%%%%%%%%%%%%%%%%%%%%%%%%%%%%%%%%%%%%%%%%%%%%%%%%%%%%%%%%%%%%
%%%%%%%%%%%%%%%%%%%%%%%%%%%%%%%%%%%%%%%%%%%%%%%%%%%%%%%%%%%%%%%%%%%%%%%%%%%%%%%%

\begin{ack}
We would like to express our gratitude to all the friends and collaborators, who contributed to the investigations
presented here, in particular to Q. B. Chen, S. Frauendorf, J. Li,  H. Z. Liang, H.
Madokoro, M. Matsuzaki, J. Peng,  P. Ring, S. Yamaji, L. F. Yu, S. Q. Zhang, Z. H. Zhang, and S. G. Zhou.
We thank R. V. F. Janssens for the careful reading of the manuscript. This work is supported by the Major State 973 Program of China (Grant No. 2013CB834400), the National Natural Science Foundation of China (Grants No. 11175002,
No. 11335002, No. 11461141002), and by U.S. Department of Energy (DOE), Office of Science, Office of Nuclear Physics, under contract DE-AC02-06CH11357.
\end{ack}

%%%%%%%%%%%%%%%%%%%%%%%%%%%%%%%%%%%%%%%%%%%%%%%%%%%%%%%%%%%%%%%%%%%%%%%%%%%%%%%%

%\bibliographystyle{mybib}
%\bibliography{paper}

\end{document}